\documentclass[runningheads]{llncs}
\usepackage[T1]{fontenc}
\usepackage{graphicx}
\usepackage{proof}
\usepackage{xcolor}
\usepackage{amssymb}
\usepackage{amsmath}
\usepackage{cmll} 
\usepackage{colonequals}
\usepackage{xspace}
\usepackage{bm}
\usepackage{hyperref}
\usepackage{thmtools,thm-restate}

\newcommand{\mypar}[1]{\smallskip\noindent{\bf #1}}

\definecolor{brickred}{rgb}{0.8, 0.25, 0.33}
\newcommand{\phl}[1]{{\color{brickred}{#1}}} 
\newcommand{\thl}[1]{#1} 

\def\bnfas{\mathrel{::=}}
\def\bnfalt{\mid}

\newcommand{\tcost}{\mathbf{cost}}
\newcommand{\tname}{\mathbf{name}}

\newcommand{\taddr}{\mathbf{addr}}

\newcommand{\msf}[1]{\mathsf{#1}}
\newcommand{\tsc}[1]{\textsc{#1}}

\newcommand*\mdelim[3]{%
	\mathopen{}\left#1%
	#3%
	\right#2\mathclose{}%
}
\newcommand*{\set}[2][]{\mdelim{\lbrace}{\rbrace_{#1}}{#2}}

\renewcommand{\emptyset}{\varnothing}
\newcommand{\lts}[1]{\xrightarrow{#1}}
\newcommand{\enc}[1]{\llcorner {#1}\lrcorner}

\newcommand{\pp}{{\ \;\boldsymbol{|}\ \;}}

\newcommand{\res}[1]{(\boldsymbol\nu #1)\,}
\newcommand{\fwd}[2]{\phl{#1\leftrightarrow #2}}
\newcommand{\wait}[2]{\phl{#1().#2}}
\newcommand{\close}[1]{\phl{{#1}[]}}
\newcommand{\send}[4]{\phl{{#1}[#2\triangleright #3].#4}}
\newcommand{\recv}[3]{\phl{{#1}(#2).#3}}
\newcommand{\Case}[3]{\phl{#1.\msf{case}(#2,#3)}}
\newcommand{\inl}[2]{\phl{{#1}[\msf{inl}].#2}}
\newcommand{\inr}[2]{\phl{{#1}[\msf{inr}].#2}}
\newcommand{\srv}[3]{\phl{\bang #1(#2).#3}}
\newcommand{\client}[3]{\phl{\query #1[#2].#3}}

\newcommand{\dual}[1]{#1^{\bot}} 
\newcommand{\one}{\mathbf{1}}
\newcommand{\tensor}{\otimes}
\newcommand{\bang}{\oc}
\newcommand{\query}{\wn}

\newcommand{\seq}{\Vdash}  
\newcommand{\cll}{\vdash}  
\newcommand{\coherence}{\vDash}  
\newcommand{\br}[2][]{[^{\phl{#1}} #2]}
\newcommand{\dbr}[2][]{[\![^{\phl{#1}} #2]\!]}
\newcommand{\pairQ}[2] { \phl{#1}:\thl{#2}}
\newcommand{\Star}{\textcolor{blue}{\ast}}
\newcommand{\Query}{\textcolor{blue}{\mathcal Q}}
\newcommand{\Left}{\textcolor{blue}{\mathcal L}} 
\newcommand{\Right}{\textcolor{blue}{\mathcal R}} 
\newcommand{\Symbol}{\textcolor{blue}{\mathcal{X}}}

\newcommand{\cut}{\tsc{Cut}\xspace}

\newdimen\mydisplayskip
\newenvironment{smallequation*}
{\par\nobreak\vskip\mydisplayskip\noindent\bgroup\small\csname equation*\endcsname}{\csname endequation*\endcsname\egroup}

\newenvironment{smallalign*}
{\par\nobreak\noindent\bgroup\small\csname align*\endcsname}{\csname endalign*\endcsname\egroup}

\newcommand\mydots{\hbox to 1em{.\hss.\hss.}}

\newcommand{\transl}[1]{\mathsf{tr}(#1)}

\begin{document}
\title{A Logical Interpretation of Asynchronous Multiparty Compatibility}
%
%
\author{Marco Carbone\inst{1}\orcidID{0000-1111-2222-3333} \and
Sonia Marin\inst{2}\orcidID{1111-2222-3333-4444} \and
Carsten Sch\"urmann\inst{1}\orcidID{2222--3333-4444-5555}}
\authorrunning{M. Carbone et al.}
%
\institute{IT University of Copenhagen, 2300 Copenhagen S, Denmark\\ 
\email{\{maca,carsten\}@itu.dk}\\
\and
School of Computer Science, University of Birmingham, B15 2TT Birmingham, United Kingdom\\
\email{s.marin@bham.ac.uk}
}
\maketitle              
\begin{abstract}
  Session types are types for specifying the protocols that
  communicating processes must follow in a concurrent system.  When
  composing two or more well-typed processes, a session typing system
  must check whether such processes are {\em multiparty compatible}, a
  property that guarantees that all sent messages are eventually
  received and no deadlock ever occurs. Previous work has shown that
  duality and the more general notion of coherence are sufficient
  syntactic conditions for guaranteeing the multiparty compatibility
  property.

  In this paper, following a propositions-as-types fashion which
  relates session types to linear logic, we generalise coherence to
  forwarders. Forwarders are processes that act as middleware by
  forwarding messages according to a given protocol. Our main result
  shows that forwarders not only generalise coherence, but fully
  capture all well-typed multiparty compatible processes.

\end{abstract}

\section{Introduction}\label{sec:introduction}
%
Session types, originally proposed by Honda et al.~\cite{HVK98}, are
type annotations that ascribe protocols to processes in a concurrent
system and determine how they behave when communicating together.
\emph{Binary session types} found a logical justification in
\emph{linear logic}, identified by Caires and
Pfenning~\cite{CP10,CP16} and later by Wadler~\cite{W12,W14}, which
establishes the following correspondences: 
session types \emph{as} linear logic propositions, processes \emph{as}
proofs, reductions in the operational semantics \emph{as} cut
reductions in linear logic, and dualilty \emph{as} a notion of
compatibility ensuring that two processes communication pattern match.
%



In 
binary session types, a sufficient condition for two protocols to be
\emph{compatible} is that their type annotations are dual: the send
action of one party must match a corresponding receive action of the
other party, and vice versa.  Because of asynchronous interleavings,
however, there are protocols that are compatible but not
dual. 
The situation is even more complex for \emph{multiparty session
  types}~\cite{HYC16}, which generalise binary session types to types
for protocols with more than two participants.
A central observation is that compatibility of sessions 
requires a property stronger than duality, ensuring that all messages
sent by any participating party will eventually be collected by
another.
In~\cite{DY13}, Deni{\'{e}}lou and Yoshida proposed the semantic
notion of \emph{multiparty
  compatibility}. 
%
The concept has then found many successful applications in the
literature~\cite{DY13,LY19,GPPSY21}.
The question whether this notion ``would be applicable to extend other
theoretical foundations such as the correspondence with linear logic
to multiparty communications'' has however not been answered since
Deni{\'{e}}lou and Yoshida's original work.


As a first step in defining a logical correspondent to multiparty
compatibility, Carbone et al.~\cite{CMSY15,CLMSW16} extended Wadler's
embedding of binary session types into classical linear logic (CLL) to
the multiparty setting by generalising logical duality to the notion
of {\em coherence}~\cite{HYC16}.
Coherence is a compatibility condition: coherent processes are
multiparty compatible, which ensures that their execution never leads
to a communication error. Coherence is characterised
proof-theoretically, and each coherence proof corresponds precisely to
a multiparty protocol specification (\emph{global type}).
An interesting observation 
is that coherence proofs
correspond to a definable subset of the processes typable in linear
logic, so-called {\em arbiters}~\cite{CLMSW16}.
%
%
%
In retrospect, the concept of coherence has sharpened our
proof-theoretic understanding of how to characterise compatibility in
multiparty session types. However, coherence, similarly to duality,
cannot capture completely the notion of {multiparty
  compatibility}. 

In this paper, we show that coherence can be generalised to a more
expressive logic of asynchronous
forwarders. 
%
\emph{Forwarders} are processes that forward messages between
endpoints according to a protocol specification.
Forwarders are more general 
than arbiters (every arbiter corresponds to a forwarder but not vice
versa) but still can be used to guide the communication of multiple
processes and guarantee they {\em communicate safely}, as done by
coherence in~\cite{CLMSW16}.
%
Our main result is that forwarders 
fully capture multiparty compatibility
which let us answer Deni{\'{e}}lou and Yoshida's original question
positively. 
In this work, we show that {\em i) any possible interleaving of a set
  of multiparty compatible processes can be represented as a
  forwarder, and, conversely, ii) if a set of processes can be
  composed by a forwarder (describing a possible execution), then such
  processes are indeed multiparty compatible.
}

%


Operationally, forwarders are a central process that decide how to
dispatch messages between peers, providing a global description of a
multiparty protocol (hence they can be seen as a generalisation of
global types). They capture the message flow by preventing messages
from being {\em duplicated}, as superfluous messages would not be
accounted for, and by preventing messages from being {\em lost},
otherwise a process might get stuck, awaiting a message. However, when
data-dependencies allow, forwarders can 
choose to receive messages from different endpoints and forward such
messages at a later point, or decide to buffer a certain number of
messages from a given sender.  Eventually, they simply re-transmit
messages after receiving them, without 
{\em computing} with them. Intuitively, this captures an interleaving
of the communications between the given endpoints.

Besides showing that they precisely capture multiparty compatibility,
this paper shows that forwarders can also replace duality or coherence
for composing well-typed processes. We achieve this logically: we
replace the linear logic cut rule with a new rule called MCutF which
allows to compose two or more processes (proofs) using a forwarder as
a condition, instead of duality~\cite{CP10} or
coherence~\cite{CLMSW16}. Our second main result is that MCutF can be
eliminated by proof reductions that correspond to asynchronous process
communications.

Although most of this paper's interest is on their logical properties,
forwarders have other interesting features related to explaining
communication patterns as they occur in practice, 
including message routing, proxy services,
and runtime monitors for message flows~\cite{JGP16}.

\smallskip

\mypar{Outline and key contributions.}  The key contributions of this
paper include
\begin{itemize}
\item a definition of {\em multiparty compatibility} for classical
  linear logic (\S~\ref{sec:mcompatibility});
\item a logical characterisation of \emph{forwarders} 
  that enjoys a sound and complete correspondence with multiparty
  compatibility (\S~\ref{sec:forwarders});

\item a composition mechanism (MCutF) for processes {\em with
    asynchronous communication} that uses forwarders instead of
  coherence, and guarantees lack of communication errors 
  (\S~\ref{sec:mcut}).
\end{itemize}
Additionally, \S~\ref{sec:preview} introduces the main concepts on an
example and \S~\ref{sec:prelim} recaps types, processes, and
CP-typing.  \S~\ref{sec:related} discusses related and future
work. Finally, concluding remarks are in \S~\ref{sec:conclusions}.

\section{Preview}\label{sec:preview}
We start with a gentle introduction to asynchronous forwarders by
informally describing the
classic \emph{2-buyer protocol}~\cite{HYC08,HYC16}, where two buyers
intend to buy a book jointly from a seller.
The first buyer 
sends the title of the book 
to the seller,
who, in turn, sends a quote to both buyers.
Then, 
the first buyer decides how much she wishes to contribute and informs
the second buyer,
who either pays the rest or cancels the transaction by informing the
seller. If the decision was made to buy the book, the second buyer will provide the seller with an address, where to ship the book to.

The three participants 
are connected through endpoints $\phl{b_1}$, $\phl{b_2}$, and $\phl s$
respectively: an endpoint acts like a socket where processes can
write/read messages to/from. Each endpoint must be used according to
its respective session type 
which gives a precise description of how each endpoint has to act.
%
For example,
$\phl{b_1}: \tname \tensor \tcost^\perp \parr \tcost \tensor \one$
says that buyer $\phl{b_1}$  first sends (expressed by $\tensor$) a value of type $\tname$
(the book title), then receives (expressed by $\parr$) a value of type $\dual\tcost$ (the
price of the book), then sends a value of type $\tcost$ (the amount of
money she wishes to contribute), and finally terminates. Here,
following classical linear logic, the annotation $\dual{}$ stands for
dual.
The behaviour of buyer $\phl{b_2}$ and seller $\phl s$ can similarly
be specified by session types, respectively
$\phl{b_2} : \tcost^\perp \parr \tcost^\perp \parr ((\taddr \tensor
\one) \oplus \one)$ and
$\phl{s} : \tname^\perp \parr \tcost \tensor \tcost\tensor
((\,\taddr\parr\perp\,)\ \&\perp)$.

Any well-typed processes 
with respective endpoints $\phl{b_1}$, $\phl{b_2}$, and~$\phl s$ are
going to execute this protocol correctly because their type
specifications are \emph{compatible}, meaning that they match.
%
%
In a multiparty setting, such compatibility can be expressed as
\emph{coherence}~\cite{CLMSW16} and more generally with the semantic
notion of \emph{multiparty compatibility}~\cite{DY13}.  Our theory of
forwarders logically captures this notion of multiparty compatibility
precisely, namely that i) anything that has been received is
eventually sent, ii) anything that is sent must have been previously
received, and iii) the order of messages between any two points must
be preserved.
%
%
%
A \emph{forwarder} captures the message flows between two or more
different endpoints. In the example above, 
a forwarder process may receive a name 
from $\phl {b_1}$, forward it to $\phl s$, and then proceed to
receiving the price 
from $\phl s$, forward it to $\phl {b_1}$ and $\phl {b_2}$,
and so on. 
%
%
A forwarder acts as a medium that, according to a prescribed protocol,
dispatches messages processes send. In the 2-buyers case, for
$\phl{P_s}$, $\phl{P_{b_1}}$, $\phl{P_{b_2}}$ respectively
implementing the seller and the two buyers, we can depict this with
the following diagram:
\begin{center}
 \includegraphics[scale=0.4]{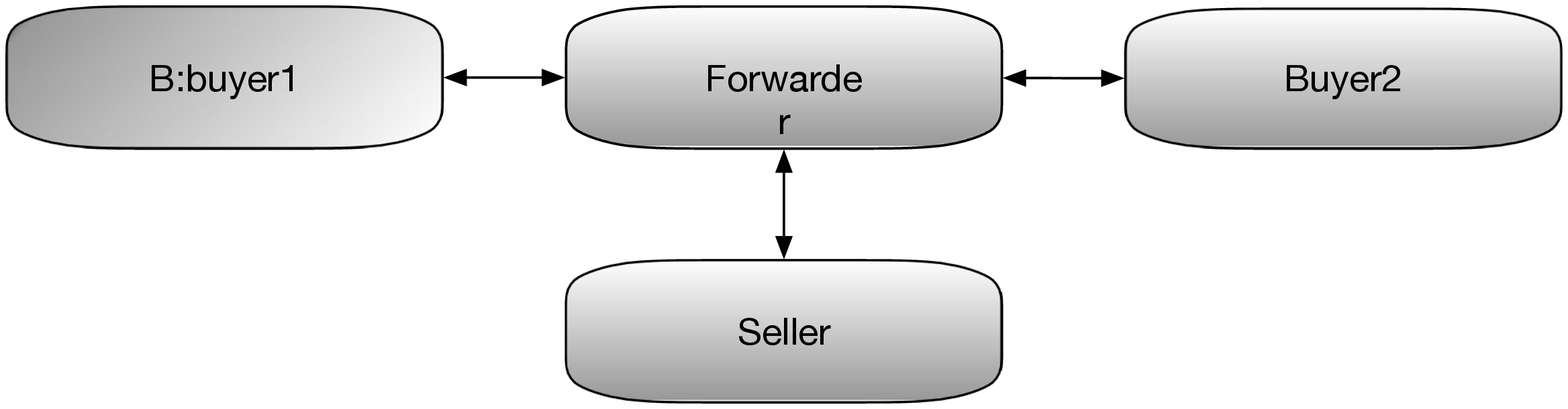}
\end{center}

In order to capture the message flows between several processes,
forwarders need to support buffering, reordering and other properties,
which are not necessarily coherent.  Consider for example two
endpoints $\phl x$ and $\phl y$ willing to communicate with the
following protocol -- called a \emph{criss-cross}: they both send a
message to each other, and then the messages are received, according
to the following types $\pairQ x{\tname\tensor \tcost \parr \one}$ and
$\pairQ y{\tcost^\perp\tensor \tname^\perp\parr \bot}$.
%
Such protocol leads to no error (assuming processes implement an
asynchronous semantics), still the two types above are not
coherent~\cite{CLMSW16} and not even dual to each other. On the other hand, we can easily write a
forwarder 
typable in the context
$\pairQ x{\tname^\perp\parr \tcost ^\perp\tensor \bot}, \pairQ
y{\tcost\parr \tname\tensor \one}$ formed by their duals, i.e., a
process that first receives on both $\phl x$ and $\phl y$ and then
forwards the received messages over to $\phl y$ and $\phl x$,
respectively.

\section{CP and Classical Linear Logic}\label{sec:prelim}
In this section,
we give an introduction to the proposition-as-sessions
approach~\cite{W14}. This comprises the syntax of types and processes
and the interpretation of processes as sequent proofs in classical
linear logic (CLL).

\mypar{Types.} Following the propositions-as-types approach, types,
taken to be propositions (formulas) of CLL, denote the way an endpoint
(a channel end) must be used at runtime. 
Their formal syntax is given by the following grammar:
\begin{equation}\label{eq:types}
	\begin{array}{lrcl}
		\mbox{Types} &
		\thl{A} &\bnfas &
	\thl{a}
	\bnfalt \thl{a^\perp}
	\bnfalt \thl{\one}
	\bnfalt \thl{\bot}
	\bnfalt \thl{A\tensor A}
	\bnfalt \thl{A\parr A}
	\bnfalt \thl{A\oplus A}
	\bnfalt \thl{A\with A}
	\bnfalt \thl{!A}
	\bnfalt \thl{?A}
	\end{array} 
\end{equation}
Atoms $a$ and negated atoms $\dual a$ are basic dual types.
Types $\one$ and $\perp$ denote an endpoint that must close with a
last synchronisation.
A type $A \tensor B$ is assigned to an endpoint that outputs a message
of type $A$ and then is used as $B$, and similarly, an endpoint of
type $A\parr B$, receives a message of type $A$ and continues as $B$.
In the situation of a branching choice, $A\oplus B$ is the type of an
endpoint that may select to go left or right and continues as $A$ or
$B$, respectively, and $A\with B$ is the type of an endpoint that
offers two choices (left or right) and then, based on such choice,
continues as $A$ or $B$.
Finally, $!A$ types an endpoint offering an unbounded number of copies
of a service of type $A$, while $?A$ types an endpoint of a client
invoking some replicated/unbounded service with behaviour $A$.

\mypar{Duality.} Operators can be grouped in pairs of duals that
reflect the input-output duality. Consequently, standard duality
$(\cdot)^\perp$ on types is inductively defined as:
\begin{smallequation*}
	\dual{(\dual{a})} = a
	\qquad \dual\one=\perp
	\qquad \dual{(A\tensor B)} = \dual A\parr\dual B
	\quad \dual {(A\oplus B)}=\dual A\with\dual B
	\quad \dual{(!A)} = ?\dual A
\end{smallequation*}
In the remainder, for any binary operators
$\oslash,\odot\in\{\tensor,\parr,\oplus,\with\}$, we sometimes write
$A\oslash B \odot C$ to mean $A \oslash (B \odot C)$.

\mypar{Processes.}  We use a standard language of {\em processes} to
represent communicating entities (including forwarders) which is a
variant of the $\pi$-calculus~\cite{MPW92} with specific communication
primitives as usually done for session calculi. Moreover, given that
the theory of this paper is based on the proposition-as-sessions
correspondence with CLL, we adopt a syntax akin to that of
Wadler~\cite{W14}. Each process can be found on the left-hand side of
the turnstyle $\vdash$ in the conclusion of each rule in
Figure~\ref{fig:cp}.
%
%
We briefly comment each process term. A link $\fwd xy$ is a binary
forwarder, i.e., a process that forwards any communication between
endpoints $\phl x$ and $\phl y$. This yields a sort of equality
relation on names: it says that endpoints $\phl x$ and $\phl y$ are
equivalent, and communicating something over $\phl x$ is like
communicating it over $\phl y$.
%
%
%
Note that we use endpoints instead of channels~\cite{V12}. The
difference is subtle: the restriction $\phl{\res {xy}}$ connects the
two endpoints $\phl x$ and $\phl y$, instead of referring to the
channel between them.
The terms $\wait xP$ and $\close x$ handle synchronisation (no message
passing); $\wait xP$ can be seen as an empty input on $\phl x$, while
$\close x$ terminates the execution of the process.
The term $\send xyPQ$ denotes a process that creates a fresh name
$\phl y$ (hence a new session), spawns a new process $\phl P$, and
then continues as $\phl Q$. The intuition behind this communication
operation is that $\phl P$ uses $\phl y$ as an interface for dealing
with the continuation of the dual primitive (denoted by term
$\recv xyR$, for some $\phl R$).
%
Note
that output messages are always fresh, as for the internal
$\pi$-calculus~\cite{S96}, hence the output term $\send xyPQ$ is a
compact version of the $\pi$-calculus term
$(\nu y)\, \overline x y.(P \,|\, Q)$.
Branching computations are handled by $\Case xPQ$, $\inl xP$ and
$\inr xP$. The former denotes a process offering two options (external
choice) from which some other process can make a selection with
$\inl xP$ or $\inr xP$ (internal choice).
Finally, $\srv xyP$ denotes a persistently available service
that can be invoked by $\client xzQ$ 
which will spawn a new session to be handled by a copy of process $\phl P$.
\begin{figure}[t]
	\begin{equation*}
		\begin{array}{c}
			\infer[\textsc{Ax}]{
				\phl{\fwd{x}{y}} 
				\cll 
				\phl x:\thl {a^\bot}, \phl y:\thl{a}
			}{}
			\qquad
			\infer[\one]{
				\phl{\close x} \cll \phl x: \thl\one
			}{} 
			\qquad
			\infer[\bot]{
				\phl {\wait xP} \cll \thl \Delta, \phl x:\thl\bot}{
				\phl P \cll \thl\Delta}
			\\[1ex]
			\infer[\tensor]{
				\phl{\send xyPQ} 
				\cll 
				\thl{\Delta_1}, \thl{\Delta_2}, \phl x:\thl{A_1 \tensor A_2}
			}{
				\phl P 
				\cll 
				\thl{\Delta_1}, \phl y:\thl A_1
				& \phl Q 
				\cll 
				\thl \Delta_2, \phl x: \thl A_2
			}
			\qquad
			\infer[\parr]
			{\phl{\recv xyP} 
				\cll 
				\thl{\Delta}, \phl x: \thl {A_1 \parr A_2}
			}       
			{\phl P \cll \thl{\Delta}, \phl y:\thl A_1, \phl x:\thl A_2}
			\\[1ex]
			     \infer[\oplus_1]
			     {\phl{\inl xP} \cll \thl\Delta, \phl x:\thl{A_1 \oplus A_2}}
			     {\phl P \cll \thl\Delta, \phl x:\thl A_1}
			     \qquad
			      \infer[\oplus_2]
			      {\phl{\inr xP} \cll \thl\Delta, \phl x:\thl{A_1 \oplus A_2}}
			      {\phl P \cll \thl\Delta, \phl x:\thl A_2}
			      \qquad
			      \infer[\with]
			      {\phl{\Case xPQ} \cll \thl \Delta, \phl x:\thl{A_1 \with A_2}}
			      {\phl P \cll \thl\Delta, \phl x:\thl A_1 
			      & 
			       \phl Q \cll \thl\Delta, \phl x:\thl A_2}
			     \\[1ex]
			     \infer[!]
			     {\phl{\srv xyP} \cll \,\thl{\query \Delta}, \phl x:\thl{\bang A}}
			     {\phl P \cll \,\thl{\query \Delta}, \phl y:\thl A}
			     \qquad
			      \infer[?]
			      {\phl{\client xyP} \cll \thl \Delta, \phl x:\thl {\query A}}
			      {\phl P \cll \thl \Delta, \phl y:\thl A}
			      \qquad
			      \infer[\textsc{w}]
			      {\phl P \cll \thl \Delta, \phl x:\thl{\query A}}
			      {\phl P \cll \thl \Delta}
				  \qquad
				  \infer[\textsc{c}]
				  {\phl P \cll \thl \Delta, \phl x:\thl{\query A}}
				  {\phl P \cll \thl \Delta, \phl x:\thl{\query A}, \phl x:\thl{\query A}}
		\end{array}
	\end{equation*}
	\caption{Sequent Calculus for CP and Classical Linear Logic} 
	\label{fig:cp}
\end{figure}

\begin{example}
  For some $\phl{P_i}$ , $\phl{Q_j}$, $ \phl{R_k}$, we provide
  possible implementations for processes $\phl{P_s}$, $\phl{P_{b_1}}$,
  and $\phl{P_{b_2}}$ from the 2-buyer example.
  \begin{displaymath}
    \begin{array}{rll}
      \phl{P_s}\ =\
      & \recv{s}{book}{}\
        \send{s}{price}{P_1}{}\
        \send s{price}{P_2}{}\
        \Case {s} {\recv s{addr}{P_3}} {P_4} \
      \\
      \phl{P_{b_1}}\ =\
      & \send{b_1}{book} {Q_1}{}\
        \recv{b_1}{price}{}\
        \send{b_1}{contr}{Q_2}{Q_3}
      \\
      \phl{P_{b_2}}\ =\
      & \recv{b_2}{price}{}\
        \recv{b_2}{contr}{}\
        \inl{b_2}{\send{b_2}{addr}{R_1}{R_2}}
    \end{array}
  \end{displaymath}
  Note that the order the two buyers receive the price is not
  relevant.
\end{example}

\mypar{CP-typing.}  As shown by Wadler~\cite{W14}, CLL proofs can be
associated to a subset of well-behaved processes, that satisfy
deadlock freedom and session fidelity.

Judgements are defined as $\phl{P} \cll \Delta$ with $\Delta$ a set of
named types, i.e.,
$\Delta \coloncolonequals \varnothing \mid \pairQ{x}{A}, \Delta$.
%
The system, called CP and reported in Figure~\ref{fig:cp}, uses CLL to
type processes.
%

CP can be extended with  a structural rule for defining composition of processes
which corresponds to the \textsc{Cut} rule from classical linear
logic:
\begin{smallequation*}
  \infer[\textsc{Cut}] {\phl{\res{xy}(P \pp Q)}
    \cll\thl\Sigma,\thl\Delta} {\phl P \cll\thl\Sigma, \phl x:\thl A &
    \phl Q \cll \thl\Delta, \phl y:\thl{\dual{A}}}
\end{smallequation*}
In linear logic this rule is admissible, i.e., 
the CLL derivations of the two premises can be combined into a derivation of the conclusion with no occurrence of the $\cut$ rule.
Moreover, this is a constructive procedure, called \emph{cut-elimination}, meaning that the proof with cut is inductively transformed into a proof without cut.
The strength of the proposition-as-type correspondence stems from the fact that it carries on to the proof level,
as it was shown that the cut-elimination steps correspond to reductions in the $\pi$-calculus~\cite{CP10,W12}.

 

%
%

\section{Multiparty Compatibility}\label{sec:mcompatibility}
Duality is a condition that guarantees composed processes (proofs) to
be well-behaved. This can be seen in rule \textsc{Cut} which composes
two processes through endpoints that are dual: cut elimination
guarantees that such processes can communicate with each other without
getting stuck or reaching an error. 
%
Multiparty compatibility~\cite{DY13,LY19,GPPSY21}, a semantic notion
that allows for the composition of multiple processes while
guaranteeing the same properties, uses session types as an abstraction
of process behaviours and simulates their execution. If no error
occurs during any such simulation then the composition is considered
compatible. 


\mypar{Extended types and queues.}  In order to define multiparty
compatibility to our logical setting with CLL formulas, we extend the
syntax of types (formulas) with annotations that make explicit where
messages should be forwarded from and to. This is similar to local
types $!{\mathsf p}.T$ and $?{\mathsf p}.T$~\cite{CDYP15} expressing
an output and an input to and from role $\mathsf p$ respectively.
%
%
%
%
The meaning of each operator and the definition of duality remains the
same as in CP.
%
\[
  \begin{array}{lrcl}
      \mbox{Types} &
      \thl{A} &\bnfas &
    \thl{a}
    \bnfalt \thl{a^\perp}
    \bnfalt \thl{\one}
    \bnfalt \thl{\bot}
    \bnfalt \thl{A\tensor A}
    \bnfalt \thl{A\parr A}
    \bnfalt \thl{A\oplus A}
    \bnfalt \thl{A\with A}
    \bnfalt \thl{!A}
    \bnfalt \thl{?A} \\
    \mbox{Local types} &
    \thl{B}  & \bnfas &
    \thl{a}
    \bnfalt \thl{a^\perp} 
    \bnfalt \thl{\one^\phl{\tilde u}}
    \bnfalt \thl{\bot^\phl{u}} 
    \bnfalt \thl{(A\tensor^\phl{\tilde u} B)} 
    \bnfalt \thl{(A\parr^\phl{u} B)} \\
    &&& \quad 
    \bnfalt  \thl{\bang^\phl{\tilde u} B}
    \bnfalt \thl{\query^\phl{u} B}      
    \bnfalt \thl{(B\oplus^\phl{u} B)}
    \bnfalt \thl{(B\with^\phl{\tilde u} B)}
  \end{array}
\]

Annotations are either single endpoints $\phl x$ or a set of endpoints
$\phl{u_1, \ldots, u_n}$. The left-hand side $A$ of $\tensor$ and
$\parr$ are not annotated (they will become dynamically labelled when
needed). 
%
We will see that units 
demonstrate some \emph{gathering} 
behaviour
%
which explains the need to annotate $\one$ 
with a non-empty list of an arbitrary number of distinct names.
We may write $\phl{\tilde u}$ for $\phl{u_1, \ldots, u_n}$ when the
size of the list is irrelevant.

Additionally to annotated types, in order to give a semantics to
types, we introduce queues defined by as
%
$
\begin{array}{rcl}
  {\Psi} & \coloncolonequals & \epsilon
  \bnfalt A\cdot \Psi
  \bnfalt {\Star}\cdot\Psi
  \bnfalt {\Left}\cdot\Psi
  \bnfalt {\Right}\cdot\Psi
  \bnfalt {\Query}\cdot\Psi
\end{array}
$.
Intuitively, a queue (FIFO) is a standard ordered list of messages. A
message can be a proposition $A$, a session termination $\Star$, a
choice $\Left$ or $\Right$, or an exponential $\Query$. Since every
ordered pair of endpoints has an associated queue, we define a queue
environment $\sigma$ as a mapping from ordered pairs of endpoints to
queues: $\sigma: (x,y)\mapsto \Psi$. In the sequel, $\sigma_\epsilon$
denotes the queue environment with only empty queues. The notation
$\sigma[(\phl x,\phl y)\mapsto \Psi]$ denotes a new environment where
the entry for $(\phl x,\phl y)$ has been updated to
$\Psi$. 
%
Finally, we are ready to define type-context semantics for an
annotated environment, i.e., an environment $\Delta$ where each
formula has been annotated (with a slight abuse of notation, we
overload the category $\Delta$):
\begin{definition}[Type-Context Semantics]\label{def:typesemantics}
  We define $\lts{\alpha}$ as the minimum relation on annotated
  $\Delta$ satisfying the following rules:
  \begin{smallequation*}
    \begin{array}{lc@{\quad}l}
      \Delta, \pairQ{x}{\perp^{\phl y}}\bullet \sigma[(\phl x,\phl y)\mapsto \Psi]
      &\lts{\phl x\perp\phl y}
      &\Delta \bullet\sigma[(\phl x,\phl y)\mapsto \Psi\cdot
        {\Star}]
      \\
      \pairQ{x}{\one^{\phl{\tilde y}}}\bullet\ \sigma_\epsilon[\{(\phl {y_i,}\phl x)\mapsto
      \Star\}_i]
      &\lts{\phl {\tilde y}\one\phl x}
      & \varnothing\bullet\sigma_\epsilon
      \\
      \pairQ{x}{\dual a}, \pairQ ya\bullet\sigma_\epsilon
      &\lts{\phl x\fwd \phl y}
      & \varnothing\bullet\sigma_\epsilon
      \\
      \Delta, \pairQ x{A \parr^{\phl y} B} \bullet \sigma[(\phl x,\phl y)\mapsto \Psi]
      &\lts{\phl x\parr\phl y}
      & \Delta, \pairQ x{B} \bullet \sigma[(\phl x,\phl y)\mapsto \Psi\cdot A]
      \\
      \Delta,
      \pairQ x{A \tensor^{\phl{\tilde y}} B}
      \bullet \sigma[\{(\phl {y_i},\phl x)\mapsto A_i\cdot\Psi_i\}_i]
      &\lts{\phl {\tilde y}\tensor \phl x[A,\{A_i\}_i]}
      &\Delta, \pairQ xB\bullet \sigma[\{(\phl {y_i},\phl x)\mapsto \Psi_i\}_i]
      \\
      \Delta, \pairQ x{A\with^{\phl{\tilde y}} B}\bullet \sigma[\{(\phl x,\phl
      {y_i})\mapsto \Psi_i\}_i]
      &\lts{\phl x\with_{\Left}\phl{\tilde y}}
      &\Delta, \pairQ x{A}\bullet \sigma[\{(\phl x,\phl {y_i})\mapsto \Psi_i\cdot{\Left}\}_i]
      \\
      \Delta, \pairQ x{A\with^{\phl{\tilde y}} B}\bullet \sigma[\{(\phl x,\phl
      {y_i})\mapsto \Psi_i\}_i]
      &\lts{\phl x\with_{\Right}\phl {\tilde y}}
      &\Delta, \pairQ x{B}\bullet \sigma[\{(\phl x,\phl {y_i})\mapsto \Psi_i\cdot{\Right}\}_i]
      \\
      \Delta, \pairQ x{A\oplus^{\phl y} B}\bullet \sigma[(\phl y,\phl x)\mapsto {\Left}\cdot\Psi]
      &\lts{\phl y\oplus_{\Left}\phl x}
      &\Delta, \pairQ x{A}\bullet \sigma[(\phl y,\phl x)\mapsto \Psi]
      \\
      \Delta, \pairQ x{A\oplus^{\phl y} B}\bullet \sigma[(\phl y,\phl x)\mapsto {\Right}\cdot
      \Psi]
      &\lts{\phl y\oplus_{\Right}\phl x}
      &\Delta, \pairQ x{B}\bullet \sigma[(\phl y,\phl x)\mapsto \Psi]
      \\
      \{\pairQ{y_i}{\query A_i}\}_i, \pairQ x{\bang^{\phl y} A}\bullet \sigma_\epsilon
      &\lts{\phl x \mathbin{!} \phl y}
      &\{\pairQ{y_i}{\query A_i}\}_i, \pairQ x{A}\bullet \sigma_\epsilon[\{(\phl x,\phl
        {y_i})\mapsto\Query\}_i]
      \\
      \Delta, \pairQ x{\query^{\phl{\tilde y}} A}\bullet \sigma[(\phl y,\phl x)\mapsto \Query\cdot \Psi]
      &\lts{\phl y \mathbin{?} \phl x}
      &\Delta, \pairQ x{A}\bullet \sigma[(\phl y,\phl x)\mapsto \Psi]
    \end{array}
  \end{smallequation*}
\end{definition}
The rules above capture an asynchronous semantics for typing contexts.
We clarify further with an example of how the semantics above
works. Assume we wish to compose three CP proofs through endpoints %
$ \Delta = \pairQ {x}{\dual A\parr B},
\pairQ {y}{\dual A\parr A\tensor C},
\pairQ {z}{A\tensor D} 
$.
The context says that $\phl x$ is receiving something of type
$\dual A$, $\phl y$ is receiving something of type $\dual A$ and then
sending something of type $A$, and, finally, $\phl z$ is sending
something of type $A$. In order to obtain an execution of $\Delta$, we
first dualise and annotate $\Delta$ (for some annotation) to, e.g.,
$ \dual \Delta = \pairQ {x}{\dual A\tensor^{\phl y} \dual B},
\pairQ {y}{A\tensor^{\phl z} \dual A\parr^{\phl x} \dual C},
\pairQ {z}{\dual A\parr^{\phl y} \dual D}
$.  Then, obtain an execution, e.g.,:
$$
\begin{array}{rl}
  \dual\Delta\bullet\sigma_\epsilon\quad
  \lts{\phl z\parr\phl y}
  & \pairQ {x}{A\tensor^{\phl y} B},
    \pairQ {y}{A\tensor^{\phl z} \dual A\parr^{\phl x} \dual C},
    \pairQ {z}{D}\ \bullet\ \sigma_\epsilon[(\phl z,\phl y)\mapsto \dual A]
  \\
  \lts{\phl z\tensor\phl y[A,\dual A]}
  & \pairQ {x}{A\tensor^{\phl y} B},
    \pairQ {y}{\dual A\parr^{\phl x} \dual C},
    \pairQ {z}{D}\ \bullet\ \sigma_\epsilon
  \\
  \lts{\phl y\parr\phl x}
  & \pairQ {x}{A\tensor^{\phl y} B},
    \pairQ {y}{\dual C},
    \pairQ {z}{D}\ \bullet\ \sigma_\epsilon[(\phl y,\phl x)\mapsto \dual A]
  \\
  \lts{\phl y\tensor\phl x[A,\dual A]}
  & \pairQ {x}{B},
    \pairQ {y}{C},
    \pairQ {z}{D} \ \bullet\ \sigma_\epsilon
\end{array}
$$

Note the general rule for the multiplicative connectors $\tensor$ and
$\parr$. In their multiparty interpretation~\cite{CLMSW16}, they
implement a gathering communication, where many $A_i\tensor B_i$ can
communicate with a single $A\parr B$. As a consequence, the $A_i$'s
are enqueued to a single endpoint which will consume such
messages. The effect of a gathering communication with such
connectives is to spawn a new session with the environment
$\{{A_i}\}_i$ shown in the label. Ideally, we could have enriched the
semantics so that it can work on different contexts running in
parallel, where $\{{A_i}\}_i$ would be added to. However, since the
need for the semantics is to define compatibility we decided to just
observe in the label. Units also have a similar gathering
behaviour. On the other hand, additives and exponentials use
broadcasting.
%
%

Using the relation on contexts above, we can define when a set of
endpoints successfully progresses without reaching an error.  This can
be formalised by the concept of live path. In the sequel, let $\alpha$
range over all the possible labels of the relation above. Moreover,
let $\alpha_1,\ldots,\alpha_n$ be a {\em path} for some annotated
$\Delta$
whenever there exist $\Delta_1,\sigma_1, \ldots\Delta_n,\sigma_n$ such
that
$\Delta\bullet\sigma_\epsilon\lts{\alpha_1}\Delta_1\bullet\sigma_1
\ldots\lts{\alpha_n}\Delta_n\bullet \sigma_n$. This path is {\em
  maximal} if there is no $\Delta_{n+1},\sigma_{n+1}, \alpha_{n+1}$
such that
$\Delta_n\bullet\sigma_n\lts{\alpha_{n+1}}\Delta_{n+1}\bullet\sigma_{n+1}$.
\begin{definition}[Live Path] A path $\tilde\alpha$ 
	for an environment $\Delta\bullet\sigma$ is {\em live} if
  $\Delta\bullet\sigma\lts{\alpha_1}
  \ldots\lts{\alpha_n}\varnothing\bullet \sigma_\epsilon$.
\end{definition}
Intuitively, a maximal path is live whenever we can consume all sends
and receives specified in the type context and all queues are empty,
i.e., an error is never reached. With this notion, we are ready to
define multiparty compatibility:
\begin{definition}[Multiparty Compatibility]\label{def:mpc}
  An environment $\Delta\bullet\sigma$ is {\em executable} if all maximal
  paths $\alpha_1,\ldots, \alpha_n$ for $\Delta$
  are live and such that
  $\alpha_i= \phl {\tilde y}\tensor \phl x[A,\set[i]{A_i}]$ implies 
  $\pairQ x\dual{A},\set[i]{\pairQ{y_i}\dual{A_i}}$ is multiparty compatible for some
  annotation.
  A context $\Delta\not=\varnothing$ is {\em multiparty compatible} if there exists an annotation such that      
  $\dual\Delta\bullet\sigma_\epsilon$ is executable.
\end{definition}
Multiparty compatibility states that all maximal paths are live, i.e.,
an error is never reached. Note that the definition above is
well-founded since propositions get smaller when reduced.

\mypar{Relationship to Previous Definitions.}
Definition~\ref{def:typesemantics} is an adaptation to CLL of the {\em
  typing environment reduction} (Definition 4.3, \cite{GPPSY21}) with
a little twist: in order not to overload notation
(cf. \S~\ref{sec:forwarders}), we are defining it on the dual of
formulas. For example, following an approach as that of
\cite{GPPSY21}, a process with an endpoint $\phl x$ of type
$A\tensor^{\phl y} B$ is meant to store something of type $A$ in the
queue from $\phl x$ to $\phl y$. In our notation, we dualise the type
of $\phl x$ to $\dual A\parr^{\phl y} \dual B$ but keep the same
behaviour, i.e., storing something of type $\dual A$ in the queue from
$\phl x$ to $\phl y$.  Moreover, for the sake of simplicity, we are
using a single queue environment $\sigma$ as a function from pairs of
endpoints to a FIFO, while Ghilezan et al.~\cite{GPPSY21} use labelled
queues attached to each endpoint of the typing context: the two
approaches are equivalent. Finally, our definition being an adaptation
to CLL, uses different language constructs. In particular, we do not
combine value passing and branching, and our $\tensor$ and $\parr$
cases spawn new sessions (hence the well-founded recursive
definition). The original definition of compatibility given by
Deni{\'{e}}lou and Yoshida is for communicating automata. Therefore,
our starting point is the adaptation to local types given by Ghilezan
et al.~\cite{GPPSY21}.



\mypar{Properties of Multiparty Compatibility.} As a consequence of
multiparty compatibility, we can formalise the lack of errors with the
following
\begin{proposition}[No Error]\label{prop:noerror}
  Let $\Delta$ be multiparty compatible and $\alpha_1,\ldots,\alpha_n$
  be a maximal path for an annotated $\dual\Delta$ such that
  $\dual\Delta\bullet\sigma_\epsilon\lts{\alpha_1}\Delta_1\bullet\sigma_1
  \ldots\lts{\alpha_n}\varnothing\bullet \sigma_\epsilon$. Then, for
  $i< n$,
  \begin{enumerate}
  \item \label{queue}
    \begin{enumerate}
    \item\label{one} 
      $\sigma_i(\phl x,\phl y)=\Star\cdot \Psi$ implies that
      $\alpha_{n} = \phl {x\tilde z}\one \phl y$;

    \item\label{two} 
      $\sigma_i(\phl x,\phl y)=A\cdot \Psi$ implies that there exists
      $k > i$ such that
      $\alpha_k = \phl {x\tilde z}\tensor \phl y[A,\{A_i\}_i]$;

    \item\label{three} 
      $\sigma_i(\phl x,\phl y)=\Left\cdot \Psi$ implies that there
      exists $k > i$ such that
      $\alpha_k = \phl {x}\oplus_{\Left} \phl y$;

    \item\label{four} 
      $\sigma_i(\phl x,\phl y)=\Right\cdot \Psi$ implies that there
      exists $k > i$ such that
      $\alpha_k = \phl {x}\oplus_{\Right} \phl y$;

    \item\label{five} 
      $\sigma_i(\phl x,\phl y)=\Query\cdot \Psi$ implies that there
      exists $k > i$ such that $\alpha_k = \phl {x} \mathbin{?} \phl y$;
    \end{enumerate}
    

  \item \label{input}
    \begin{enumerate}
    \item\label{seven} 
      $\Delta_i = \Delta_i', \pairQ x{\one^{\phl{\tilde y}} }$, then
      $\alpha_n=\phl {\tilde y}\one \phl x$;

    \item\label{eight} 
      $\Delta_i = \Delta_i', \pairQ x{A\tensor^{\phl{\tilde y}} B}$, then there exists
      $k> i$ such that
      $\alpha_k = \phl {\tilde y}\tensor \phl x[\{A_i\}_i]$;

    \item\label{nine} 
      $\Delta_i = \Delta_i', \pairQ x{A\oplus^{\phl y} B}$, then there exists
      $k> i$ such that $\alpha_k = \phl y\oplus_{\Left} \phl x$ or
      $\alpha_k = \phl y\oplus_{\Right} \phl x$;

    \item\label{ten} 
      $\Delta_i = \Delta_i', \pairQ x{\query^{\phl y} A}$, then there
      exists $k> i$ such that
      $\alpha_k = \phl {y} \mathbin{?} \phl {x}$.
    \end{enumerate}
  \end{enumerate}
\end{proposition}

Conditions in \eqref{queue} state that every message that has been
enqueued is eventually consumed.
On the other hand, conditions in~\eqref{input} state that every input
instruction is eventually executed.
Note also that since we have no infinite computations, we do not need
to consider fairness.

\mypar{Are annotations important?} A careful reader may be wondering
why the definitions of type-context semantics and multiparty
compatibility are not given for annotation-free
contexts. Unfortunately, doing so would make multiparty compatibility
too strong since we would have to allow for messages to be sent to
different endpoints in different paths.
As an example, $ \pairQ {x}{A\tensor \dual B},
\pairQ {y}{A\tensor \dual A\parr \dual C},
\pairQ {z}{\dual A\parr\dual D}
$ can get stuck if $\phl z$ communicates with $\phl x$ first,
violating property \eqref{eight} in Proposition~\ref{prop:noerror}.
Note that previous definitions of multiparty compatibility for
multiparty session types~\cite{LY19,GPPSY21} do indeed use
annotations.

%
%
 
\section{Asynchronous Forwarders}\label{sec:forwarders}
%
%
Forwarders form a subclass of processes with some special features but
that are also typable in classical linear logic.
I.e., our goal is to identify all those CP processes that are also
forwarders. In order to do so, we must add further information in the
standard CP contexts.

\mypar{Contexts.}
What we need is to be able to enforce the main features that
characterise a forwarder, namely i) anything received must be
forwarded, ii) anything that is going to be sent must be something
that has been previously received, and iii) the order of such messages
between any two points must be preserved.
%
%
In order to enforce these
requirements, we add more information to the standard CP
judgement. For example, let us consider the input process $\recv
xyP$. In CP, the typing environment for such process must be such that
endpoint $\phl x$ has type $A\parr B$ such that $\phl P$ has type
$\pairQ yA$ and $\pairQ xB$. However, the context is not telling us at
all that $\phl y$ is actually a message that has been received and, as
such, it should not be used by $\phl P$ for further communications but
just forwarded over some other channel. In order to remember this fact
when we type the subprocess $\phl P$, we actually insert $\pairQ yA$
into a queue that belongs to endpoint $\phl x$ where we put all the
types of messages received over it. I.e., when typing $\phl P$, the
context will contain $\dbr{\Psi}\br[u]{\pairQ yA}\pairQ xB$. That
still means that $\phl x$ must have type $B$ and $\phl y$ must have
type $A$ in $\phl P$, but also that $\pairQ yA$ has been received over
$\phl x$ (it is in $\phl x$'s queue) and we are intending to forward
it to endpoint $\phl u$. Moreover, $\Psi$ contains the types of
messages that have been previously received over $\phl x$.
The forwarders behave asynchronously. 
They can input arbitrarily many messages, which are enqueued at the arrival point, without blocking the possibility of producing an output from the same endpoint.
This behaviour is captured by the notion of queues of \emph{boxed}
messages, i.e.~messages that are in-transit.
\begin{equation*}
  \begin{array}{rcllllllll}
	\dbr {\Psi} & \coloncolonequals & \varnothing 
	& \mid\quad \br[u]{\Star}\dbr \Psi
	& \mid\quad \br[u]{\pairQ yA}\dbr \Psi
	 & \mid\quad \br[u]{\Query}\dbr \Psi
	 & \mid\quad \br[u]{\Left}\dbr \Psi
	 & \mid\quad \br[u]{\Right}\dbr \Psi
\end{array}
\end{equation*}
A queue element $\br[u]{\pairQ xA}$ expresses that $\phl x$ of type
$\thl A$ has been received and will need to later be forwarded to
endpoint $\phl u$.
Similarly, $\br[u]{\Star}$ indicates that a request for closing a
session has been received and must be forwarded to 
$\phl u$. $\br[u]{\Left}$ (or $\br[u]{\Right}$) and $\br[u]{\Query}$
indicate that 
a branching request and server invocation, respectively, has been
received and must be forwarded. 

%
%
The order of messages needing to be forwarded to \emph{independent}
endpoints is irrelevant.
Hence, we consider queue
$\dbr{\Psi_1}\br[x]{\ldots}\br[y]{\ldots}\dbr{\Psi_2}$ equivalent to
queue $ \dbr{\Psi_1}\br[y]{\ldots}\br[x]{\ldots}\dbr{\Psi_2}$
whenever $\phl x\not= \phl y$.
For a given endpoint $\phl x$ however the order of two messages
$\br[x]{\ldots}\br[x]{\ldots}$ is crucial and must be maintained
throughout the forwarding. This follows the idea of having a queue for
every ordered pair of endpoints in the type-context semantics in
Definition~\ref{def:typesemantics}.
By attaching a queue to each endpoint we get a typing context
\begin{equation*}
\begin{array}{rcllllllll}
  \Gamma & \coloncolonequals
  & \quad\varnothing
  & \quad\mid\quad \Gamma, \dbr \Psi \pairQ xB
  & \quad\mid\quad \Gamma, \dbr\Psi \pairQ x\cdot 
\end{array}
\end{equation*}
The element $\dbr{\Psi}\pairQ xB$ of a context $\Gamma$ indicates that
the messages in $\dbr{\Psi}$ have been received at endpoint $\phl x$.
The special case $\dbr{\Psi}\pairQ x\cdot$ is denoting the situation
when endpoint $\phl x$ no longer needs to be used for communication,
but still has a non-empty queue of messages to forward.

When forwarding to many endpoints, we use $\br[\tilde u]{\Symbol}$ for
denoting $\br[u_1]{\Symbol}\ldots\br[u_n]{\Symbol}$, with
$\phl{\tilde u} = \phl{u_1}, \ldots, \phl{u_n}$.
In this case, we also assume the implicit rewriting
$\br[\emptyset]{\Symbol}\dbr{\Psi} \equiv \dbr{\Psi}$.

\mypar{Judgements and rules.}
A judgement denoted by $\phl P \seq \Gamma$ types the forwarder
processes $\phl P$ that connects the endpoints in $\Gamma$.
The rules enforce the asynchronous forwarding behaviour by adding
elements to queues using 
rules for $\bot$ and $\parr$, which forces them to be later removed
from queues by the corresponding 
rules for $\one$ and $\tensor$.  
The rules are reported in Fig.~\ref{fig:mult-rules}.

\begin{figure}[t]
\begin{displaymath}\small
	\begin{array}{c}
		\infer[\tsc{Ax}]{
			\fwd xy\seq\pairQ{x}{\dual a}, \pairQ ya
		}{}
		\qquad
		\infer[\bot]{
			\wait xP \seq \Gamma, \dbr{\Psi}\pairQ{x}{\bot^\phl{u}}
		}{
			\phl{P} \seq \Gamma, \dbr{\Psi}\br[u]{\Star}\pairQ{x}{\cdot}
		}
		\qquad	
		\infer[\one\ (\tilde u \not= \emptyset)]{
			\close x \seq \set[i]{\br[x]{\Star}\pairQ{u_i}{\cdot}}, \pairQ{x}{\one^\phl{\tilde u}}
		}{ }
		\\\\
		\infer[\parr]{
			\phl{\recv xyP} 
			\seq 
			\Gamma, \dbr{\Psi}\pairQ x{A \parr^\phl{u} B} 
		}{
			\phl P 
			\seq 
			\Gamma, \dbr{\Psi}\br[u]{\pairQ yA}\pairQ xB
		}
		\qquad
		\infer[\tensor\ (\tilde u \not= \emptyset)]{
			\phl{\send xy{P}Q} 
			\seq 
			\Gamma,
			\big\{\br[\phl x]{\pairQ{y_i}{A_i}}\dbr{\Psi_i}\pairQ{u_i}{C_i}\big\}_i,
			\dbr{\Psi}\pairQ x{A \tensor^\phl{\tilde u} B}
		}{
			\phl {P}
			\seq
			\big\{\pairQ{y_i}{A_i}\big\}_i,\pairQ yA
			& 
			\phl Q 
			\seq
			\Gamma, \big\{\dbr{\Psi_i}\pairQ{u_i}{A_i}\big\}_i,
			\dbr{\Psi}\pairQ xB
		}
		\\\\
		\infer[\with\ (\tilde u
          \not= \emptyset)]{
			\Case{x}{P}{Q}
			\seq \Gamma, \dbr{\Psi}\pairQ x{A \with^\phl{\tilde u
          } B}
		}{
			\phl P\seq \Gamma, \dbr{\Psi}\br[\tilde u]{\Left}\pairQ xA
			&
			\phl Q\seq \Gamma, \dbr{\Psi}\br[\tilde u
                   ]{\Right}\pairQ xB
		}
		\\\\
		\infer[\oplus_l]{
			\inl xP\seq \Gamma, \br[x]{\Left}\dbr{\Psi_z} \phl z: \thl C,
			\dbr{\Psi_x}\pairQ x{A \oplus^\phl{z} B}
		}{
			\phl P\seq \Gamma, \dbr{\Psi_z} \phl z: \thl C, \dbr{\Psi_x}\pairQ xA
		}
		\qquad
		\infer[\oplus_r]{
			\inr xP\seq \Gamma, \br[x]{\Right}\dbr{\Psi_z} \phl z: \thl C,
			\dbr{\Psi_x}\pairQ x{A \oplus^\phl{z} B}
		}{
			\phl P\seq \Gamma, \dbr{\Psi_z} \phl z: \thl C, \dbr{\Psi_x}\pairQ xB
		}
		\\\\
		\infer[!\ (\tilde u \not= \emptyset)]{
			\srv xyP \seq \set[i]{\pairQ{u_i}{\query B_i}}, \pairQ x{\bang^\phl{\tilde u} A} 
		}{
			\phl{P} \seq \set[i]{\pairQ{u_i}{\query B_i}}, \br[\tilde u]{\Query}\pairQ y A
		}
		\qquad
		\infer[?]{
			\client xyP \seq \Gamma, \br[x]{\Query}\dbr{\Psi_z}\pairQ zC, \dbr{\Psi_x}\pairQ x{\query^\phl{z} A}
		}{
			\phl{P}\seq \Gamma, \dbr{\Psi_z}\pairQ zC, \dbr{\Psi_x}\pairQ yA
		}
	\end{array}
\end{displaymath} 
\caption{Proof System for Forwarders}
\label{fig:mult-rules}
\end{figure}

Rule \tsc{Ax} is identical to the one of CP.
Rules $\one$ and $\bot$ forward a request to close a session. Rule $\bot$ receives the request on endpoint $\phl x$ and enqueues it as $\br[u]{\Star}$ if it needs to forward it to $\phl u$. Note that in the premiss of $\bot$ the endpoint is terminated pending the remaining messages in the corresponding queue being dispatched. Eventually all endpoints but one will be terminated in the same manner. 
Rule $\one$ will then be applicable. Note that the behaviour of $\wait xP$ and $\close x$ work as gathering, 
several terminated endpoints connect to the last active endpoint typed with a $\one$.
%
Rules $\tensor$ and $\parr$ forward a message. Rule $\parr$ receives the message $\pairQ y A$ and enqueues it as $\br[u]{\pairQ y A}$ to be forwarded to endpoint $\phl u$. 
Dually, rule $\tensor$ applied to a $\tensor$ sends the messages at
the top of the queues of endpoints $\phl {u_i}$'s, meaning that
several messages are sent at the same time. Messages will be picked
from queues belonging to distinct endpoints, as a consequence, the
left premiss of $\tensor$ rule spawns a new forwarder consisting of
the gathered messages.

In the case of additives and exponentials, the behaviour is actually
\emph{broadcasting}, that is, an external choice $\br[u]{\Left}$ or
$\br[u]{\Right}$, or a server opening $\br[u]{\Query}$, resp., is
received and can be used several (at least one) times to guide
internal choices or server requests, resp., later on.

Note how annotations put constraints on how the proof is constructed,
e.g., annotating $\pairQ x{A\parr B}$ with $\phl u$ ensures us that
the proof will contain a $\tensor$-rule application on endpoint
$\phl u$. 

\begin{example}[Multiplicative Criss-cross]\label{ex:criss-cross}
  $\phl P \colonequals \recv xu{\recv yv{\send y{u'}{\fwd u{u'}}{\send
        x{v'}{\fwd {v'}v}{\wait x{\close y}}}}}$ is one of the
  forwarders that can prove the compatibility of the types involved in
  the criss-cross protocol (in \S~\ref{sec:preview}), as illustrated
  by the derivation below.
	
	\begin{smallequation*}
	\infer[\parr]{
		\phl P
		\seq
		\pairQ x{\tname^\perp\parr^\phl{y} \tcost ^\perp\tensor^\phl{y} \bot^\phl{y}},
		\pairQ y{\tcost\parr^\phl{x}  \tname\tensor^\phl{x} \one^\phl{x}}
		}{
		\infer[\parr]{
			\recv yv{\send y{u'}{F_1}{\send x{v'}{F_2}{F_3}}}
			\seq
			\br[y]{\pairQ{u}{\tname^\perp}}\pairQ x{ \tcost ^\perp\tensor^\phl{y} \bot^\phl{y}},
			\pairQ y{\tcost\parr^\phl{x}  \tname\tensor^\phl{x} \one^\phl{x}}
			}{
			\infer[\tensor]{
				\send y{u'}{F_1}{\send x{v'}{F_2}{F_3}}
				\seq
				\br[y]{\pairQ{u}{\tname^\perp}}\pairQ x{ \tcost ^\perp\tensor ^\phl{y}\bot^\phl{y}},
				\br[x]{\pairQ{v}{\tcost}}\pairQ y{\tname\tensor^\phl{x} \one^\phl{x}}
				}{
				\infer[\tsc{Ax}]{
					\seq
					\pairQ{u}{\tname^\perp}, \pairQ{u'}{\tname}
					}{
					\phl{F_1} \colonequals \fwd u{u'}
					}
					&
					\infer[\tensor]{
						\send x{v'}{F_2}{F_3}
						\seq 
						\pairQ x{ \tcost ^\perp\tensor ^\phl{y}\bot^\phl{y}},
						\br[x]{\pairQ{v}{\tcost}}\pairQ y{\one^\phl{x}}
						}{
						\infer[\tsc{Ax}]{
							\seq
							\pairQ{v'}{\tcost ^\perp}, \pairQ{v}{\tcost}
							}{
							\phl{F_2} \colonequals \fwd {v'}v
							}
						&
						\infer[\bot]{
							\phl{F_3}
							\seq 
							\pairQ x{\bot^\phl{y}},
							\pairQ y{\one^\phl{x}}
							}{
							\infer[\one]{
								\close y
								\seq 
								\br[y]{\Star}\pairQ x{\cdot},
								\pairQ y{\one}
								}{
								\phl{F_3} \colonequals \wait x{\close y}
								}
							}
						}
				}
			}
		}
	\end{smallequation*}
\end{example}


      \begin{example}
        The following is a forwarder for the 2-buyer protocol, for
        some $\phl {T_i}$'s.
  \begin{displaymath}
    \begin{array}{l}
      \recv{b_1'}{book} {}\
      \send{s'}{book}{T_1}{}\
      \recv{s'}{price}{}\
      \recv {s'}{price}{}\
      \send{b_1'}{price}{T_2}{}\
      \send{b_2'}{price}{T_3}{}\\
      \recv{b_1'}{contr}{}\
      \send{b_2'}{contr}{T_4}{}\quad
      \Case{b_2'}{
      \inl {s'}{}\
      \recv{b_2'}{addr}{}\
      \send s{addr}{T_5}{\ T_6} \quad}
      {\quad T_7}\
    \end{array}
  \end{displaymath}
      
\end{example}


%

\mypar{Properties of Forwarders.}  We write $\enc{B}$ for the formula
obtained from any $B$ by removing all the annotations.
We state that every forwarder is also a CP process, the
embedding $\enc{\cdot}$ being extended to contexts and queus as:
\begin{smallequation*}
	\begin{array}{l@{\qquad\qquad}l@{\qquad}l}
		\enc{\dbr{\Psi}\pairQ{x}{B}, \Gamma} = \enc{\dbr{\Psi}}, \pairQ{x}{\enc B}, \enc{\Gamma}
		&
		\enc{\dbr{\Psi}\pairQ{x}{\cdot}, \Gamma} = \enc{\dbr{\Psi}}, \enc{\Gamma}
		\\[1ex]
		\enc{\br[u]{\pairQ{y}{A}}\dbr{\Psi}}  = \pairQ{y}{A} , \enc{\dbr{\Psi}}
		&
		\enc{\br[u]{\Symbol}\dbr{\Psi}}  = \enc{\br[u]{\Star}\dbr{\Psi}} = \enc{\dbr{\Psi}}
		\\
		&\text{where $\Symbol \in \set{\Left, \Right, \Query}$}
	\end{array}
\end{smallequation*}

\begin{proposition}\label{prop:embed}
	Any forwarder is typable in CP, i.e., if $\phl P \seq \Gamma$, then  $\phl P \cll \enc{\Gamma}$.
\end{proposition}

Moreover, forwarders enjoy an invertibility property, i.e., all its
rules are invertible.
In CLL, the rules $\tensor$ or $\oplus$ are not invertible because of the choice involved either in splitting the context in the conclusion of $\tensor$ into the two premisses or the choice of either disjuncts for $\oplus$. 
In our case on the other hand, the annotations put
extra syntactic constraints on what can be derived and hence are restricting these choices to a unique one and as a result the rules are invertible.
This is formalised by the following.

\begin{proposition}\label{prop:invertibility}
  All the forwarder rules are invertible, that is, for any rule
	if there exists a forwarder $\phl F$ such that $\phl F \seq \Gamma$, the conclusion of the rule, 
	there is a forwarder $\phl{F_i} \seq \Gamma_i$, for each of its premiss, $i= 1$ or $2$. 
\end{proposition}

\begin{proof}
  For each rule, the proof is standard and follows by induction on $\phl F$.
  For example, for $\oplus_l$, let us suppose that 
  $\phl F  \seq \Gamma, \br[x]{\Left}\dbr{\Psi_z} \phl z: \thl C,
  \dbr{\Psi_x}\pairQ x{A \oplus^\phl{z} B}$.
  If $\phl F = \inl{x}{F_1}$, then we directly get that
  $\phl{F_1} \seq \Gamma, \dbr{\Psi_z} \phl z: \thl C,
  \dbr{\Psi_x}\pairQ xA$.
	
  Otherwise, if $\phl F$ starts with any other process operator, we
  apply an inductive step.
  E.g., let $\phl F = \recv uv{F'}$, meaning that
  $\Gamma = \Gamma', \dbr{\Psi_u}\pairQ u{A \parr^\phl{w} B}$ and that
  $\phl{F'} \seq \Gamma', \dbr{\Psi_u}\br[w]{\pairQ vA}\pairQ uB,
  \br[x]{\Left}\dbr{\Psi_z} \phl z: \thl C, \dbr{\Psi_x}\pairQ x{A
    \oplus^\phl{z} B}$.
  By induction hypothesis (given that $\phl{F'}$ is a subprocess of
  $\phl F$), we know there exists $\phl{F_1'}$ such that
  $\phl{F_1'} \seq \Gamma', \dbr{\Psi_u}\br[w]{\pairQ vA}\pairQ uB,
  \dbr{\Psi_z} \phl z: \thl C, \dbr{\Psi_x}\pairQ xA$.
  From there we can derive $\phl{F_1} = \recv uv{F_1'}$.
  The rest of the cases follow in a similar way.
\end{proof}

\mypar{A Note on Compositionality and Cut Elimination.} The 
sequent calculus we presented enjoys cut elimination. That means that
forwarders can be composed, and their composition is still a
forwarder. The way we compose forwarders follows a standard cut rule
augmented with extra machinery to deal with annotations. Additionally,
the cut elimination proof provides a semantics for forwarders, in the
proposition-as-types style.
For space reasons, we have not included its technical development, but
it can be found in our extended note~\cite{CMS21b}.

\mypar{Relation to Multiparty Compatibility.}
Forwarders relate to transitions in the type-context semantics
introduced in the previous section.
%
In order to formalise this, we first give a translation from
type-contexts into forwarder contexts: 
\begin{itemize}
	\item$\transl{\varnothing\bullet\sigma_\epsilon} = \varnothing$.
	
	\item 	$\transl{\Delta, \pairQ{x}{A} \bullet \sigma\cup\set[i]{(\phl {y_i,}\phl x)\mapsto\Psi_i} }
	\colonequals 
	\dbr[y_1]{\Psi_1}\ldots\dbr[y_n]{\Psi_n}\pairQ{x}{A}, \transl{\Delta\bullet\sigma}$
	for a type environment $\Delta = \set[i]{\pairQ{y_i}{\Psi_i}}$ and a queue environment $\sigma$ mapping the endpoints $\phl{y_i}$;
\end{itemize}
We use the extended notation $\dbr[u]{\Psi}$ to signify that all the brackets in $\dbr{\Psi}$ are labelled by $\phl u$.
%
%
%

\begin{lemma}\label{lem:delta-to-gamma}
	Let $\Delta\bullet\sigma$ be a type-context and $\Gamma= \transl{\Delta\bullet\sigma}$.
	\begin{enumerate}
        \item\label{item:rule} if there exists $\alpha$ and
          $\Delta'\bullet\sigma'$ such that
          $\Delta\bullet\sigma \lts{\alpha} \Delta'\bullet\sigma'$
          then there exists a rule from forwarders such that $\Gamma$
          is an instance of its conclusion and
          $\Gamma'= \transl{\Delta'\bullet\sigma'}$ is an instance of
          (one of) the premiss(es);
		\item\label{item:no-rule} otherwise, either $\Delta =\varnothing$ and $\sigma = \sigma_\epsilon$ or there is no forwarder $\phl F$ such that $\phl F \seq \Gamma$.
	\end{enumerate}
\end{lemma}

\begin{proof}
	Follows from a simple inspection of the transitions in Def.~\ref{def:typesemantics} and the rules in Fig.~\ref{fig:mult-rules}.
\end{proof}

Thus, we can conclude this section by proving that forwarders are
actually characterising multiparty compatibility for CP processes
(processes that are well-typed in CLL).

\begin{restatable}{theorem}{correspondence}
	\label{thm:correspondence}
$\Delta$ is multiparty compatible iff there exists a forwarder
$\phl F$ such that $\phl F\seq\dual\Delta$ where each connective in $\dual\Delta$ is annotated.
\end{restatable}


\begin{proof}[Proof Sketch]
	See Appendix~\ref{app:correspondence} for the full proof.
	From left to right, we need to prove more generally that if $\Delta\bullet\sigma$ is executable, then there exists a forwarder
	$\phl F$ such that $\phl F\seq\transl{\Delta\bullet\sigma}$,
	by induction on the size of $\Delta$, defined as the sum of the formula sizes in $\Delta$.
	From right to left can be proven by contrapositive, using
        Lemma~\ref{lem:delta-to-gamma} and
        Proposition~\ref{prop:invertibility}.
\end{proof}

\section{Composing Processes with Asynchronous
  Forwarders}\label{sec:mcut}
In this section, we show how to use forwarders for logically composing
CP processes.

\mypar{Multiparty Process Composition.}  We start by focusing on the
{\em structural} rule that can be added to CP, namely the
$\cut$, 
as seen in Section~\ref{sec:prelim}.  
%
Rule \tsc{Cut} corresponds to parallel composition of processes. The
implicit side condition that this rule uses is \emph{duality}, i.e.,
we can compose two processes if endpoints $\phl x$ and $\phl y$ have a
dual type. Carbone et al.~\cite{CLMSW16} generalise the concept of
duality
to that of {\em coherence}. Coherence,
denoted by $\coherence$, generalises duality to many endpoints,
allowing for a cut rule that composes many processes in parallel 
\begin{smallequation*}
  \infer[\tsc{MCut}]
  {\phl{\res{\tilde x:G} (R_1 \pp \ldots \pp R_n)} \cll \set[i\le n]{\Sigma_i}}
  {
    \set[i\le n]{\phl {R_i} \cll \thl\Sigma_i, \pairQ{x_i}{A_i}}
    &
    \phl G \coherence \thl\set[i\le n]{\pairQ{x_i}{A_i}}}
\end{smallequation*}
The judgement $\phl G \coherence \set[i\le n]{\pairQ{x_i}{A_i}}$ intuitively
says that the $\pairQ{x_i}{A_i}$'s are compatible and the execution of the
$\phl{R_i}$ will proceed without any error.
Such a result is formalised by an \tsc{MCut} elimination theorem
analogous to the one of CP. We leave $\phl G$ abstract here: it is a
proof term and it corresponds to a global type (see~\cite{CLMSW16}).

Our goal here is to replace the notion of coherence with an
asynchronous forwarder $\phl Q$, yielding the rule
\begin{smallequation*}
  \infer[\tsc{MCutF}]
  {\phl{\res{\tilde x:Q}(R_1 \pp \ldots \pp R_n)} \cll \set[i\le n]{\Sigma_i}}
  {
    \set[i\le n]{\phl {R_i} \cll \thl\Sigma_i, \pairQ{x_i}{A_i}}
    &
    \phl Q \seq \thl\set[i\le n]{\pairQ{x_i}\dual{A_i}}
  }
\end{smallequation*}
Asynchronous forwarders are more general than coherence: every
coherence proof can be transformed into an \emph{arbiter}
process~\cite{CLMSW16}, which is indeed a forwarder, while there are
judgements that are not coherent but are provable in our forwarders
(see Example~\ref{ex:criss-cross}).
In the rule \tsc{MCutF},
the role of a forwarder (replacing coherence) is to be a
middleware that decides whom to forward messages to. This means that
when a process $\phl{R_i}$ sends a message to the middleware, the 
message must be stored by the forwarder, who will later
forward it to the right receiver.
Since our goal is to show that \tsc{MCutF} is admissible (and hence we
can eliminate it from any correct proof), we extend such rule to
account for messages in transit that are temporarily held by the
forwarder. In order to do so, we use the forwarders queues and some
extra premises and define $\tsc{MCutQ}$ as:
\begin{smallequation*}
  \infer[]{
  	\phl{\res{\tilde x: Q \br{\tilde y \lhd P_1, \ldots, P_m}} (R_1 \pp \ldots \pp R_n)} \cll \set[j\le m]{\Delta_j}, \set[i\le n]{\Sigma_i}
  }{
  	\set[j\le m]{\phl {P_j} \cll \thl\Delta_j, \pairQ{y_j}{A_j}}
    &
    \set[i\le n]{\phl {R_i} \cll \thl\Sigma_i, \pairQ{x_i}{B_i}}
    &
    \phl Q \seq \set[i\le n]{\dbr{\Psi_i}\pairQ{x_i}\dual{B_i}}, \set[n < i \le p]{ \dbr{\Psi_i}\pairQ{x_i}{\cdot}}
  }
\end{smallequation*}
We have three types of process terms:
$\phl{P_j}$'s, $\phl{R_i}$'s and $\phl{Q}$. Processes $\phl{R_i}$'s
are the processes that we are composing, implementing a multiparty
session. $\phl Q$ is the forwarder whose role is to certify
compatibility and to determine, at run time, who talks to
whom. Finally, processes $\phl{P_i}$'s must be linked to messages in the
forwarder queue. 
Such processes are there because of the way $\tensor$
and $\parr$ work in linear logic. This will become clearer when we
look at the reduction steps that lead to cut admissibility. 
This imposes a side condition on the rule, namely that 
\begin{smallequation*}
	\bigcup_{i\le p}{\Psi_i} \setminus
\set{\Star} = \set[j\le 
m]{\pairQ{y_j}{\dual{A_j}}}
\end{smallequation*}
Note that we need to introduce a new syntax for this new structural
rule: in the process
$\phl{\res{\tilde x: Q \br{\tilde y \lhd P_1, \ldots, P_m}} (R_1 \pp
  \ldots \pp R_n)}$, the list $\phl{P_1, \ldots, P_m}$ denotes those
messages (processes) in transit that are going to form a new session
after the communication has taken place.
In the remainder we (slightly abusively) abbreviate both  $\phl{\set{P_1, \ldots, P_m}}$ and $\phl{(R_1 \pp \ldots \pp R_n)}$ as $\phl{\tilde P}$ and $\phl{\tilde R}$ respectively.

\mypar{Semantics and \tsc{MCutF}-admissibility.} 
We now formally show that \tsc{MCutF} is admissible, yielding a
semantics for our extended CP (with \tsc{MCutF}) in a
proposition-as-types fashion.
In order to do so,
we consider some cases from the multiplicative fragment (see appendix
for all cases).
In the sequel, 
$\Gamma = \set[i\le n]{\dbr{\Psi_i}\pairQ{x_i}\dual{B_i}}, \set[n < i
\le p]{ \dbr{\Psi_i}\pairQ{x_i}{\cdot}}$ and
$\Gamma{-k}= \Gamma \setminus
\set{\dbr{\Psi_k}\pairQ{x_k}\dual{B_k}}$.
Also, we omit (indicated as ``$\ldots$'') the premises of the
$\tsc{MCutQ}$ that do not play a role in the reduction at hand, and
assume that they are always the same as above, that is,
$\set[j\le m]{\phl {P_j} \cll \thl\Delta_j, \pairQ{y_j}{A_j}}$ and
$\set[i\le n]{\phl {R_i} \cll \thl\Sigma_i, \pairQ{x_i}{B_i}}$.

\noindent {\it Send Message ($\tensor$).} This is the case when a
process intends to send a message, which corresponds to a
$\tensor$ rule. As a consequence, the forwarder has to be ready to
receive the message (to then forward it later):
\begin{smallequation*}
  \infer[\tsc{MCutQ}]{
  	\phl{\res{x\tilde x:\recv xyQ\br{{\tilde y} \lhd\tilde P}}(\send{x}{y}{P}{R} \pp \tilde R)} \cll \Delta, \Sigma, \set[j\le m]{\Delta_j}, \set[i\le n]{\Sigma_i}
  }
  {
    \infer[\tensor]
    {
      \phl{\send{x}{y}{P}{R}}\cll \Delta,\Sigma, \pairQ x{A\tensor B}
    }
    {
      \phl P\cll \Delta,\pairQ yA
      &
      \phl R\cll \Sigma,\pairQ xB
    }
    & \quad\ldots\quad
    &
    \infer[\parr]
    {
      \phl{\recv xyQ} \seq 
      \dbr{\Psi}\pairQ x{\dual A\parr^{\phl{x_k}}\dual B}, 
      \Gamma
    }
    {
      \phl Q\seq \dbr{\Psi}\br[x_k]{\pairQ y{\dual A}}\pairQ x{\dual B}, 
      \Gamma
    }
  }
\end{smallequation*}
The process on the left is ready to send the message to the forwarder.
By the annotation on the forwarder, it follows that the message will
have to be forwarded to endpoint $\phl{x_k}$, at a later
stage. Observe that the nature of $\tensor$ forces us to deal with the
process $\phl P$: the idea is that when the forwarder will finalise
the communication (by sending to a process $\phl{R'}$ owning endpoint
$\phl{x_k}$) process $\phl P$ will be composed with $\phl {R'}$. For
now, we obtain the reductum:
\begin{smallequation*}
  \infer[\tsc{MCutQ}]{
  	\phl{\res{x\tilde x:Q\br{{y,\tilde y} \lhd P, \tilde P}}(R \pp \tilde R)} \cll \Delta, \Sigma, \set[j\le m]{\Delta_j}, \set[i\le n]{\Sigma_i}
  }
  {
    	\phl P\cll\Delta,\pairQ yA
    	&
    	\phl R\cll\Sigma,\pairQ xB
    & \quad\ldots\quad
    &
    \phl Q\seq 
    \dbr{\Psi}\br[x_k]{\pairQ y\dual{A}}\pairQ x\dual{B}, 
    \Gamma
  }
\end{smallequation*}

\noindent {\it Receive Message ($\parr$).} At a later point,
the forwarder will be able to complete the forwarding operation by
connecting with a process ready to receive ($\parr$ rule):
\begin{smallequation*}
  \infer[\tsc{}]
  {\phl{\res{x\tilde x:\send xy{S}Q\br{z,\tilde y \lhd P,\tilde P}}(\recv{x}{y}R \pp \tilde R)}
  	\cll 
  	\Delta, \Sigma, \set[j\le m]{\Delta_j}, \set[i\le n]{\Sigma_i}
  }
  {
    \begin{array}{c}
      \phl{P} \cll \Delta, \pairQ{z}\dual{A}
  \hspace*{2cm}
  \ldots
  \hspace*{2cm}
    \\ 
    \infer[\parr]
    {
    	\phl{\recv{x}{y}{R}}\cll \Sigma, \pairQ x{A\parr B}
    }
    {
    	\phl{R}\cll \Sigma, \pairQ yA, \pairQ x{B}
    }
	\quad
    \infer[\tensor]{
    	\phl{\send xy{S}Q} 
    	\seq 
    	\dbr{\Psi_x}\pairQ x{\dual A \tensor^{\phl{x_k}} \dual B}, \br[x]{\pairQ{z}{A}}\dbr{\Psi_k}\pairQ{x_k}\dual{B_k},
    	\Gamma-k
    }{
    	\phl S
    	\seq
    	\pairQ{z}{A}, \pairQ y{\dual A}
    	&
    	\phl Q
    	\seq
    	\dbr{\Psi_x}\pairQ x{\dual B}, 
    	\Gamma
    }
    \end{array}
  }
\end{smallequation*}
Key ingredients are process $\phl P$ with endpoint $\phl z$
of type $\dual A$, endpoint $\phl{x_k}$ in the forwarder with a
boxed endpoint $\phl z$ with type $A$, and process
$\phl{\recv{x}{y}{R}}$ ready to receive. 

After reduction, we obtain the following:
\begin{smallequation*}
  \infer[\tsc{MCutQ}]
  {\phl{\res{x\tilde x:Q\br{\tilde y \lhd \tilde P}}(\res{yz:S}(R\pp P) \pp \tilde R)} \cll 
  \Delta, \Sigma, \set[j\le m]{\Delta_j}, \set[i\le n]{\Sigma_i}
	}
  {
      \phl{\res{yz:S}(R\pp P)}\cll\Sigma, \Delta, \pairQ xB
    %
    &\quad\ldots\quad
    &
    \phl Q
    \seq
    \dbr{\Psi_x}\pairQ x{\dual B}, 
    \Gamma
  }
\end{smallequation*}
Where the left premiss is obtained as follows:
\begin{smallequation*}
	    \infer[\tsc{MCutQ}]
	{
		\phl{\res{yz:S}(R\pp P)}\cll\Sigma, \Delta, \pairQ xB
	}
	{
		\phl R\cll \Sigma, \pairQ yA, \pairQ x{B}
		&
		\phl P\cll\Delta,\pairQ z{\dual A}
		&
		\phl S\seq\pairQ zA,\pairQ y{\dual A}
	}
\end{smallequation*}
meaning that now the message (namely process $\phl P$) has finally been
delivered and it can be directly linked to $\phl R$ with a new (but smaller) \tsc{MCutQ}.

These reductions (the full set can be found in
Appendix~\ref{app:generalise}) allow us to prove the key lemma of this
section.
\begin{lemma}[$\tsc{MCutQ}$ Admissibility]
  If
  $\set[j\le m]{\phl {P_j} \cll \thl\Delta_j, \pairQ{y_j}{A_j}}$
  and
  
  \noindent
  $\set[i\le n]{\phl {R_i} \cll \thl\Sigma_i, \pairQ{x_i}{B_i}}$
  and 
  $\phl Q \seq \set[i\le n]{\dbr{\Psi_i}\pairQ{x_i}\dual{B_i}}, \set[n < i \le p]{ \dbr{\Psi_i}\pairQ{x_i}{\cdot}}$
  then there exists a process $\phl S$ such that   
  $\phl{\res{\tilde x: Q \br{\tilde y \lhd \tilde P}} \tilde R} \Rightarrow^* \phl S$
  and
  $\phl S \cll \set[j\le m]{\Delta_j}, \set[i\le n]{\Sigma_i}$.
\end{lemma}
\begin{proof}[Proof Sketch]
  By lexicographic induction on ($i$) the sum of sizes of the $B_i$'s
  and ($ii$) the sum of sizes of the $\phl{R_i}$'s.
  Some of the key and base cases have been detailed above; the others
  can be found in Appendix~\ref{app:generalise}.
  The commutative cases are straightforward and only need to consider
  the possible last rule applied to a premiss of the form
  $\phl {R_i} \cll \thl\Sigma_i, \pairQ{x_i}{B_i}$.
\end{proof}

We can finally conclude with the following theorem as a special case.

\begin{restatable}[$\tsc{MCutF}$ Admissibility]{theorem}{cutadm}
	\label{thm:cutadm}
	If
	$\set[i\le n]{\phl {R_i} \cll \thl\Sigma_i, \pairQ{x_i}{A_i}}$
	and
	$\phl Q \seq \thl\set[i\le n]{\pairQ{x_i}\dual{A_i}}$
	then there exists a process $\phl S$ such that   
	$\phl{\res{\tilde x:Q}(R_1 \pp \ldots \pp R_n)}\Rightarrow^* \phl S$
	and
	$\phl S \cll \set[j\le m]{\Delta_j}, \set[i\le n]{\Sigma_i}$.
\end{restatable}

%
%

\section{Related Work}\label{sec:related}
Our work takes~\cite{CLMSW16} as a starting point.
Guided by CLL, we set out to explore if coherence can be broken down
into more elementary logical rules which led us to introduce
forwarders. As a result, forwarders provide a more general notion of
compatibility.
An earlier unpublished version of this work~\cite{CMS21} proposes
synchronous forwarders, i.e., the restriction of forwarders with only
buffers of size one.
In that case, we show that we can always construct a coherence proof
from a synchronous forwarder.  However, synchronous forwarders fail to
capture all the possible interleaving of an arbiter (encoding of
coherence to processes).

Caires and Perez~\cite{CP16} also study multiparty session types in
the context of intuitionistic linear logic by translating global types
to processes, called \emph{mediums}.
Their work does not start from a logical account of global types
(their global types are just syntactic terms). But, as previous
work~\cite{CLMSW16}, they do generate arbiters as linear logic proofs,
which are special instances of forwarders. In this work, we generalise
this approach to characterise exactly which processes can justify the
compatibility of several processes. In a more recent work, van den
Heuvel and P{\'{e}}rez~\cite{HP22} use {\em routers} in order to
provide a decentralised analysis of multiparty protocols. Routers act
as point-to-point forwarders but their types, called {\em relative
  types}, carry extra information on causality of events that are not
local. As in~\cite{CP16}, this approach is confined to global types
and is therefore not complete wrt multiparty compatibility.
%

%

Another logical interpretation of multiparty compatibility is proposed
by Horne~\cite{horne2020session}.
It uses the additonal expressivity of BV, a generalisation of CLL with
a non-commutative sequential operator, but only considers a fragment
that is expressible in the sequent calculus, unlike their previous
work which relied on the full strength of deep
inference~\cite{ciobanu2015behavioural}.
It also allows one to consider compatible processes beyond duality but
only for simply typed processes, which cannot spawn other processes.
The main advantage of this approach is the fact that annotations are
not needed (but can be recovered in the typing).

Sangiorgi~\cite{S96}, probably the first to treat forwarders for the
$\pi$-calculus, uses binary forwarders, i.e., processes that only
forward between two channels, which are equivalent to our $\fwd xy$. We
attribute our result to the line of work that originated in 2010 by
Caires and Pfenning~\cite{CP10},
where forwarders \emph{\`a la Sangiorgi} were introduced as processes
to be typed by the axiom rule in linear logic. Van den Heuvel and
Perez~\cite{Heuvel2020SessionTS} have recently developed a version of
linear logic that encompasses both classical and intuitionistic logic,
presenting a unified view on binary forwarders in both logics.

Gardner et al.~\cite{GLW07} study the expressivity of the linear
forwarder calculus, by encoding the asynchronous $\pi$-calculus (since
it can encode distributed choice).  The linear forwarder calculus is a
variant of the (asynchronous) $\pi$-calculus that has binary
forwarders and a restriction on the input $x(y).P$ such that $y$
cannot be used for communicating (but only for forwarding). Such a
restriction is similar to the intuition behind our forwarders,
with the key difference that their methodology would not apply to some of our
session-based primitives. 

Barbanera and Dezani~\cite{BD19} study multiparty session types as
\emph{gateways} that work as a medium among many interacting parties,
forwarding communications between two multiparty sessions. Such
mechanism reminds us of our forwarder composition: indeed, in their
related work discussion they do mention that their gateways could be
modelled by 
a ``connection-cut''.

Recent work~\cite{JGP16,GJP18} proposes a variant of linear logic that
models {\em identity providers}, monitors that are like our forwarders
but restricted to between two channels. Identity providers are
asynchronous, i.e., they allow unbounded buffering of messages before
forwarding. Our forwarders can be seen as a generalisation to
multiparty monitors. Multiparty monitors are also addressed by Hamers
and Jongmans~\cite{HJ20}, but not in a linear logic context.

Our forwarding mechanism may be confused with that of
locality~\cite{MS04}, which is addressed logically by Caires et
al.~\cite{CPT16}. Locality requires that received channels cannot be
used for inputs (which must occur at the location where the channel
was created). 
In our case, we do not allow received channels to be used at all until
a new forwarder is spawned.

\section{Conclusions and Future Work}\label{sec:conclusions}
We showed that forwarders are a logical characterisation of multiparty
compatibility and they can safely replace coherence for composing all
well-typed compatible communicating processes.
%
%
Below, we discuss some aspects of forwarders 
and identify possible future
extensions. 



\mypar{Improving Multiparty Compatibility?} Multiparty compatibility
concerns the error-free composition of processes that communicate by
enqueueing/dequeueing messages into/from pair-wise distinct FIFO
queues.  This work is not about improving multiparty compatibility,
unlike, e.g., \cite{GPPSY21}. In this paper, we take the definition of
multiparty compatibility and give it a logical characterisation, in
the spirit of the research line started in~\cite{CP10}. Forwarders are
derived from our logical characterisation, this is our novel
contribution.

\mypar{Are Forwarders Centralised?} Following the approach taken for
arbiters~\cite{CLMSW16} and mediums~\cite{CP16}, forwarders provide an
orchestration of the message flows between the composed processes. In
order to step to a fully decentralised setting, it is necessary to
redefine rule MCut such that i) queues are no longer embedded in
forwarders and ii) annotations in the forwarders are transferred to
the composed processes. This two steps are immediate and the
correctness of this follows, also immediately, by
Theorem~\ref{thm:correspondence}, since the type-context semantics in
Definition~\ref{def:typesemantics} is indeed fully decentralised. Note
that a similar decentralisation approach is also done for coherence in
\cite{CLMSW16}.


\mypar{Cut Elimination for Forwarders.} An unpublished longer version
of this work~\cite{CMS21b} addresses forwarder composition through a
CUT rule for forwarders. Due to annotations, such rule requires some
subtle technical details for which there was no space in this
paper. This CUT rule for forwarders is proven to be admissible through
an algorithm of CUT elimination which mimics how messages are
transferred from one forwarder to another.

\mypar{Process Language.} Our process language is based Wadler's
CP~\cite{W14}, without polymorphic communications.  
We conjecture that our forwarders can be smoothly extended to
polymorphic types $\exists X.A$ and $\forall X.A$.
As future work, we plan to consider a further extension to support
recursion in the style of Toninho et al.~\cite{TCP14}. That will
require an extended notion of multiparty compatibility dealing with
infinite paths as done by Ghilezan et al.~\cite{GPPSY21}.

\mypar{Variants of Linear Logic.} In this paper, we have chosen to
base our theory on CLL for two main reasons.  Coherence is indeed
defined by Carbone et al.~\cite{CLMSW16} in terms of CLL and therefore
our results can immediately be related to theirs without further
investigations. 
%
An earlier version of forwarders was based on intuitionistic linear
logic, but moving to CLL required fewer rules and greatly improved the
presentation.  Nevertheless, our results should be easily reproducible
in intuitionistic linear logic.
A different approach could be to include non-commutative operators
which could encode our FIFO queues, e.g., using the work on
non-commutative subexponentials by Kanovich et al.~\cite{KKNS18}. We
leave this as future work.

\mypar{Beyond Linear Logic.}
Another interesting avenue would be to understand how the queueing mechanism of forwarders can be treated within the graphical proof system of~\cite{acclavio2022graphical}.
Indeed, they observed that queues of length greater than 3 could not be expressed as linear logic formulas so they designed a proof system that works not only on LL formulas (or even its generalisation with a sequential operator called BV) but on more general graphs.


\mypar{Variants of Coherence.} Our results show that forwarders are a
generalisation of coherence proofs.  Indeed, coherence would
correspond to the notion of \emph{synchronous
  forwarders}~\cite{CMS21}, the restriction of forwarders with only
buffers of size one.  As a follow-up, we would like to investigate,
whether other syntactic restrictions of forwarders also induce
interesting generalised notions of coherence, and, as a consequence,
generalisations of global types.

%
%

\subsubsection{Acknowledgements} We would like to thank Frank Pfenning
for initially suggesting the idea of characterising arbiters as linear
logic proofs, and Nobuko Yoshida and Ross Horne for stimulating
discussions.

%
%
%
\bibliographystyle{splncs04}
\bibliography{biblio}

\newpage
\appendix

%
%

\section{Compatiblity vs. Forwarders}\label{app:correspondence}
\correspondence*

\begin{proof} 
	
	Left to right: 
	We will prove a slightly more general statement, namely that if $\Delta\bullet\sigma$ is executable, then there exists a forwarder
	$\phl F$ such that $\phl F\seq\transl{\Delta\bullet\sigma}$.
	
	Assume that $\Delta\bullet\sigma$ is executable, i.e., by Def.~\ref{def:mpc},
	all maximal paths $\alpha_1,\ldots, \alpha_n$ for $\Delta\bullet\sigma$ are live 
	and such that
	$\alpha_i= \phl {\tilde y}\tensor \phl x[A,\{A_i\}_i]$ implies
	$\pairQ x\dual{A},\{\pairQ{y_i}\dual{A_i}\}_i$ is multiparty compatible (i.e., its dual is executable on $\sigma_\epsilon$ for some
	annotation).

	We reason by induction on the size of $\Delta$, defined as the sum of the formula sizes in $\Delta$.
	As $\Delta \not= \varnothing$ and any maximal path for $\Delta\bullet\sigma$ is live, there must exist $\alpha_1$ such that $\Delta\bullet\sigma \lts{\alpha_1} \Delta_1\bullet\sigma_1$. 
	%
	By Lemma~\ref{lem:delta-to-gamma}.\ref{item:rule}, there exists a rule $R_{\alpha_1}$ such that $\Gamma = \transl{\Delta\bullet\sigma}$ is an instance of the conclusion of $R_{\alpha_1}$ and $\Gamma_1 = \transl{\Delta_1\bullet\sigma_1}$ is an instance of (one of) the premiss(es) of $R_{\alpha_1}$.
	
	In the base case where $\Delta_1 = \varnothing$ and $\sigma_1 = \sigma_\epsilon$, we must have
	$\alpha_1 = \phl {\tilde y}\one\phl x$ or $\phl x\fwd \phl y$ and the associated rule $R_{\alpha_1}$(i.e., $\one$ or $\textsc{Ax}$ resp.) does not have any premisses giving us resp.~$\close{x} \seq \Gamma$ or $\fwd{x}{y} \seq\Gamma$.
	
	If the rule $R_{\alpha_1}$ has a single premiss, then by induction there exists $\phl{F_1}\seq \Gamma_1$ and by applying the process operator that corresponds to $R_{\alpha_1}$ we obtain the appropriate $\phl F\seq\Gamma$.
	%
	
	If the rule $R_{\alpha_1}$ is branching, i.e., $\tensor$ or $\with$, we have to treat each case separately:
	\begin{itemize}
		\item If $\alpha_1 = \phl {\tilde y}\tensor \phl x[A,\{A_i\}_i]$, 
		then $R_{\alpha_1}$ is the $\tensor$-rule and $\Gamma_1$ is its right premiss.
		Its left premiss would consist of $\pairQ{z}{A}, \set[i]{\pairQ{z_i}{A_i}}$ for some fresh endpoint names $\phl z, \phl{\tilde z}$.
		We know that $\pairQ{z}{\dual A}, \set[i]{\pairQ{z_i}\dual{A_i}}$ form a multiparty compatible environment, and since $\mathsf{size}(\pairQ{z}{ A}, \set[i]{\pairQ{z_i}{A_i}}) < \mathsf{\Delta}$, we have by induction that there exists a forwarder $\phl{F_0} \seq  \pairQ{z}{ A}, \set[i]{\pairQ{z_i}{A_i}}$.
		Hence, $\send{x}{z}{F_0}{F_1} \seq \Gamma$.

		\item If $\alpha_1 = \phl x\with_{\Left}\phl {\tilde y}$, then $R_{\alpha_1}$ is the $\with$-rule and $\Gamma_1$ is its left premiss. 
		However, this means that the transition $\alpha_2 = \phl x\with_{\Right}\phl {\tilde y}$ was also possible as $\Delta\bullet\sigma \lts{\alpha_2}\Delta_2\bullet\sigma_2$. 
		By Lemma~\ref{lem:delta-to-gamma}.\ref{item:rule}, we get a rule $R_{\alpha_2}$ which is also a $\with$-rule and of which $\Gamma_2 = \transl{\Delta_2\bullet \sigma_2}$ is the right premiss. 
		By induction applied to $\Delta_1\bullet\sigma_1$ we get $\phl{F_1}\seq\Gamma_1$ and by induction applied to $\Delta_2\bullet\sigma_2$ we get $\phl{F_2}\seq\Gamma_2$.
		Hence, $\Case{x}{F_1}{F_2} \seq \Gamma$.
	\end{itemize}

	Right to left: 
	By contrapositive, suppose that $\Delta$ is not multiparty compatible. 
	There is then a maximal path $\alpha_1,\ldots, \alpha_n$ for any annotation of $\dual\Delta$ (or for one of its sub-environments, in which case the reasoning is easily adapted) which is not live.
	Fix an annotation on $\dual\Delta$, there is $\dual\Delta\bullet\sigma_\epsilon\lts{\alpha_1}\ldots \lts{\alpha_n} \Delta_n\bullet\sigma_n$ such that there is no available transition from $\Delta_n\bullet\sigma_n$ despite $\Delta_n \not= \varnothing$ or $\sigma_n \not= \sigma_\epsilon$.
	By Lemma~\ref{lem:delta-to-gamma}.\ref{item:rule}, we can
        construct a branch of a derivation in forwarders starting from
        root $\dual\Delta$, following $\tilde\alpha$ step-by-step and
        endinn at the leaf on $\Gamma$ = translation of
        $\Delta_n\bullet\sigma_n$.
	By Lemma~\ref{lem:delta-to-gamma}.\ref{item:no-rule}, $\Gamma$
        is not provable in our proof system for forwarders, i.e.,
        there is no $\phl F$ such that $\phl F \seq \Gamma$.
	By Lemma~\ref{prop:invertibility} this means that there is no correct derivation of $\dual\Delta$ (as otherwise applying rules bottom-up would preserve this property all the way to the leaf $\Gamma$). 
\end{proof}

\section{MCutF admissibility}\label{app:generalise}
In this appendix, we display all cases for \tsc{MCutF} elimination
that we need to prove:

\cutadm*

\begin{displaymath}\small
  \infer[\tsc{MCutF}]{
  	\phl{\res{\tilde x: \res{\tilde y}Q \br{\tilde P}} \tilde R} \cll \tilde\Delta, \tilde\Sigma
  }{
  	\set[j\le m]{\phl {P_j} \cll \thl\Delta_i, \pairQ{y_j}{A_j}}
    &
    \set[i\le n]{\phl {R_i} \cll \thl\Sigma_i, \pairQ{x_i}{B_i}}
    &
    \phl Q \seq \set[i\le n]{\dbr{\Psi_i}\pairQ{x_i}\dual{B_i}}, \set[n < i \le p]{ \dbr{\Psi_i}\pairQ{x_i}{\cdot}}
  }
\end{displaymath}
The rule has a developed side condition: $\bigcup_{i\le p}{\Psi_i} \setminus
\set{\Left, \Right, \Query, \Star} = \set[j\le
m]{\pairQ{y_j}{\dual{A_j}}}$.

\mypar {Send Message ($\tensor$).}
\begin{displaymath}\small
  \infer[\tsc{MCutF}]{
  	\phl{\res{x\tilde x:\res{\tilde y}\recv xyQ\br{\tilde P}}(\send{x}{y}{P}{R} \pp \tilde R)} \cll \Delta, \Sigma, \tilde\Delta, \tilde\Sigma
  }
  {
    \infer[\tensor]
    {
      \phl{\send{x}{y}{P}{R}}\cll \Delta,\Sigma, \pairQ x{A\tensor B}
    }
    {
      \phl P\cll \Delta,\pairQ yA
      &
      \phl R\cll \Sigma,\pairQ xB
    }
    & \quad\ldots\quad
    &
    \infer[\parr]
    {
      \phl{\recv xyQ} \seq 
      \dbr{\Psi}\pairQ x{\dual A\parr^{\phl{x_k}}\dual B}, 
      \Gamma
    }
    {
      \phl Q\seq \dbr{\Psi}\br[x_k]{\pairQ y{\dual A}}\pairQ x{\dual B}, 
      \Gamma
    }
  }
\end{displaymath}
\qquad\qquad$\Longrightarrow$ 
\begin{displaymath}\small
  \infer[\tsc{MCutF}]{
  	\phl{\res{x\tilde x:\res{y\tilde y}Q\br{P, \tilde P}}(R \pp \tilde R)} \cll \Delta, \Sigma, \tilde\Delta, \tilde\Sigma
  }
  {
    	\phl P\cll\Delta,\pairQ yA
    	&
    	\phl R\cll\Sigma,\pairQ xB
    & \quad\ldots\quad
    &
    \phl Q\seq 
    \dbr{\Psi}\br[x_k]{\pairQ y\dual{A}}\pairQ x\dual{B}, 
    \Gamma
  }
\end{displaymath}

\mypar{Receive Message ($\parr$).} 
\begin{displaymath}\small
  \infer[\tsc{MCutF}]
  {\phl{\res{x\tilde x:\res{z\tilde y}\send xy{S}Q\br{P,\tilde P}}(\recv{x}{y}R \pp \tilde R)}
  	\cll 
  	\Delta, \Sigma, \tilde\Delta, \tilde\Sigma
  }
  {
    \begin{array}{ll}
      \phl{P} \cll \Delta, \pairQ{z}\dual{A}
      \\[2mm]
      \infer[\parr]
      {
      \phl{\recv{x}{y}{R}}\cll \Sigma, \pairQ x{A\parr B}
      }
      {
      \phl{R}\cll \Sigma, \pairQ yA, \pairQ x{B}
      }
    \end{array}
    &
    \ldots
    &
    \infer[\tensor]{
      \phl{\send xy{S}Q} 
      \seq 
      \dbr{\Psi_x}\pairQ x{\dual A \tensor^{\phl{x_k}} \dual B}, \br[x]{\pairQ{z}{A}}\dbr{\Psi_k}\pairQ{x_k}\dual{B_k},
      \Gamma-k
    }{
    	\phl S
    	\seq
    	\pairQ{z}{A}, \pairQ y{\dual A}
    	&
      \phl Q
      \seq
	\dbr{\Psi_x}\pairQ x{\dual B}, 
	\Gamma
    }
  }
\end{displaymath}
\qquad\qquad$\Longrightarrow$ 
\begin{displaymath}\small
  \infer[\tsc{MCutF}]
  {\phl{\res{x\tilde x:\res{\tilde y}Q\br{\tilde P}}(\res{yz:S}(R\pp P) \pp \tilde R)} \cll 
  \Delta, \Sigma, \tilde\Delta, \tilde\Sigma
	}
  {
    \infer[\tsc{MCutF}]
    {
      \phl{\res{yz:S}(R\pp P)}\seq\Sigma, \Delta, \pairQ xB
    }
    {
      \phl R\cll \Sigma, \pairQ yA, \pairQ x{B}
      &
      \phl P\cll\Delta,\pairQ z{\dual A}
      &
      \phl S\seq\pairQ zA,\pairQ y{\dual A}
    }
    &\quad\ldots\quad
    &
    \phl Q
    \seq
    \dbr{\Psi_x}\pairQ x{\dual B}, 
    \Gamma
  }
\end{displaymath}
Note that now the message (namely process $\phl P$) has finally been
delivered and it can be directly linked with a new \tsc{MCutF}.

\mypar{Internal choice ($\oplus$).} 
\begin{displaymath}\small
	\infer[\tsc{MCutF}]
	{\phl{\res{x\tilde x:\res{\tilde y}\Case x{Q}S\br{\tilde P}}(\inl{x}{R} \pp \tilde R)}
		\cll 
		\Sigma, \tilde\Delta, \tilde\Sigma
	}
	{
		\infer[\oplus_l]
		{
			\phl{\inl{x}{R}}\cll \Sigma, \pairQ x{A\oplus B}
		}
		{
			\phl{R}\cll \Sigma, \pairQ x A
		}
		&\quad\ldots\quad
		&
		\infer[\with]{
			\phl{\Case x{Q}S} 
			\seq
			\dbr{\Psi_x}\pairQ x{\dual A \with^{\phl{x_k}} \dual B}, 
			\Gamma
		}{
			\phl Q
			\seq
			\dbr{\Psi_x}\br[x_k]{\Left}\pairQ x{\dual A}, 
			\Gamma
			&
			\phl S
			\seq
			\dbr{\Psi_x}\br[x_k]{\Right}\pairQ x{\dual B}, 
			\Gamma
		}
	}
\end{displaymath}
\qquad\qquad$\Longrightarrow$ 
\begin{displaymath}\small
	\infer[\tsc{MCutF}]
	{\phl{\res{x\tilde x:\res{\tilde y}Q\br{\tilde P}}(R \pp \tilde R)}
		\cll 
		\Sigma, \tilde\Delta, \tilde\Sigma
	}
	{
		\phl{R}\cll \Sigma, \pairQ x A
		&\quad\ldots\quad
		&
			\phl Q
			\seq
			\dbr{\Psi_x}\br[x_k]{\Left}\pairQ x{\dual A}, 
			\Gamma
	}
\end{displaymath}

\mypar{External choice ($\with$).} 
\begin{displaymath}\small
	\infer[\tsc{MCutF}]{
		\phl{\res{x\tilde x:\res{\tilde y}\inl{x}{Q}\br{\tilde P}}(\Case{x}{R}{S} \pp \tilde R)} \cll \Sigma, \tilde\Delta, \tilde\Sigma
	}
	{
		\infer[\with]
		{
			\phl{\Case{x}{R}{S}}\cll \Sigma,\pairQ x{A\with B}
		}
		{
			\phl R\cll \Sigma,\pairQ yA
			&
			\phl S\cll \Sigma,\pairQ xB
		}
		& \quad\ldots\quad
		&
		\infer[\oplus_l]{
			\inl xQ\seq 
			\dbr{\Psi_x}\pairQ x{\dual A \oplus^{x_k} \dual B}, \br[x]{\Left}\dbr{\Psi_k} \phl {x_k}: \thl \dual{B_k},
			\Gamma-k
		}{
		\phl Q\seq 
			\dbr{\Psi_x}\pairQ x{\dual A}, 
			\Gamma
		}
	}
\end{displaymath}
\qquad\qquad$\Longrightarrow$
\begin{displaymath}\small
	\infer[\tsc{MCutF}]{
		\phl{\res{x\tilde x:\res{\tilde y}{Q}\br{\tilde P}}({R} \pp \tilde R)} \cll \Sigma, \tilde\Delta, \tilde\Sigma
	}
	{
			\phl R\cll \Sigma,\pairQ yA
		&\quad\ldots\quad
		&
			\phl Q\seq 
			\dbr{\Psi_x}\pairQ x{\dual A}, 
			\Gamma
	}
\end{displaymath}

\mypar{Axiom.} 
%
\begin{displaymath}\small
	\infer[\tsc{MCutF}]{
		\phl{\res{x_1x_2:\fwd {x_1}{x_2}}(\fwd {x_1}{z_1} \pp \fwd {x_2}{z_2})} \cll \pairQ {z_1}{\dual a}, \pairQ {z_2}{a}
	}
	{
		\infer[\tsc{Ax}]{
			\fwd {x_1}{z_1}\cll\pairQ{x_1}{a}, \pairQ {z_1}{\dual a}
		}{}
		&
		\infer[\tsc{Ax}]{
			\fwd {x_2}{z_2} \cll\pairQ{x_2}{\dual a}, \pairQ {z_2}{a}
		}{}
		&
		\infer[\tsc{Ax}]{
			\fwd {x_1}{x_2} \seq\pairQ{x_1}{\dual a}, \pairQ {x_2}{a}
		}{}
	}
\end{displaymath}
\qquad\qquad$\Longrightarrow$
\begin{displaymath}\small
	\infer[\tsc{Ax}]{
		\fwd {z_1}{z_2}\cll\pairQ {z_1}{\dual a}, \pairQ {z_2}{a}
	}{}
\end{displaymath}

\mypar{Close ($\one$).}
\begin{displaymath}\small
	\infer[\tsc{MCutF}]{
		\phl{\res{x\tilde x:\res{\tilde y}\wait xQ\br{\tilde P}}(\close x \pp \tilde R)} \cll \tilde\Delta, \tilde\Sigma
	}
	{
		\infer[\one]
		{\phl{\close x} \cll \phl x: \thl\one}
		{ }
		&\quad\ldots\quad
		&
		\infer[\bot]{
			\wait xQ \seq 
			\dbr{\Psi_x}\pairQ{x}{\bot^{x_k}}, 
			\Gamma
		}{
			\phl{Q} \seq 
			\dbr{\Psi_x}\br[x_k]{\Star}\pairQ{x}{\cdot}, 
			\Gamma
		}
	}
\end{displaymath}
\qquad\qquad$\Longrightarrow$
\begin{displaymath}\small
	\infer[\tsc{MCutF}]{
		\phl{\res{x\tilde x:\res{\tilde y}Q\br{\tilde P}}(\tilde R)} \cll \tilde\Delta, \tilde\Sigma
	}
	{
		\quad\ldots\quad
		&
			\phl{Q} \seq 
			\dbr{\Psi_x}\br[x_k]{\Star}\pairQ{x}{\cdot}, 
			\Gamma
	}
\end{displaymath}

\mypar{Wait ($\bot$).}
\begin{displaymath}\small
	\infer[\tsc{MCutF}]{
		\phl{\res{x:\close x }(\wait xP)} \cll \Delta
	}
	{
		\infer[\bot]
		{\phl {\wait xP} \cll \thl \Delta, \phl x:\thl\bot}
		{\phl P \cll \thl\Delta}
		&
		\infer[\one\ (\vec u \not= \emptyset)]{
			\close x \seq \pairQ{x}{\one^{\tilde x}}, \set[i]{\br[x]{\Star}\pairQ{x_i}{\cdot}}
		}{ }
	}
\end{displaymath}
\qquad\qquad$\Longrightarrow$\quad
${\phl P \cll \thl\Delta}$

\mypar{Server query ($\bang$).} 
\begin{displaymath}\small
	\infer[\tsc{MCutF}]{
		\phl{\res{x\tilde x:\res{\tilde y}\client xzQ\br{\tilde P}}(\srv xzR \pp \tilde R)} \cll \query\Sigma, \tilde\Delta, \tilde\Sigma
	}
	{
		\infer[!]{
			\phl{\srv xzR} \cll \,\thl{\query \Sigma}, \phl x:\thl{\bang A}
			}{
			\phl R \cll \,\thl{\query \Sigma}, \phl z:\thl A
			}
		& \quad\ldots\quad
		&
		\infer[?]{
			\client xzQ \seq \Gamma-k, \br[x]{\Query}\dbr{\Psi_k}\pairQ {x_k}{\dual{B_k}}, \dbr{\Psi_x}\pairQ x{\query^\phl{x_k} \dual{A}}
		}{
			\phl{Q}\seq \Gamma, \dbr{\Psi_x}\pairQ z\dual{A}
		}
	}
\end{displaymath}
\qquad\qquad$\Longrightarrow$
\begin{displaymath}\small
	\infer[\tsc{MCutF}]{
		\phl{\res{z\tilde x:\res{\tilde y}{Q}\br{\tilde P}}({R} \pp \tilde R)} \cll \query\Sigma, \tilde\Delta, \tilde\Sigma
	}
	{
			\phl R \cll \,\thl{\query \Sigma}, \phl z:\thl A
		&\quad\ldots\quad
		&
			\phl{Q}\seq \Gamma, \dbr{\Psi_x}\pairQ z\dual{A}
	}
\end{displaymath}

\mypar{Server opening ($\query$).} 
\begin{displaymath}\small
	\infer[\tsc{MCutF}]
	{\phl{\res{x\tilde x:\res{\tilde y}\srv xzQ\br{\tilde P}}(\client xzR \pp \tilde R)} 
		\cll 
		\Sigma,\tilde\Sigma
	}
	{
		\infer[?]
		{\phl{\client xzR} \cll \thl \Sigma, \phl x:\thl {\query A}}
		{\phl R \cll \thl \Sigma, \phl z:\thl A}
		&\quad\ldots\quad
		&
		\infer[!\ (\tilde u \not= \emptyset)]{
			\srv xzQ \seq \set[i]{\pairQ{x_i}{\query \dual{A_i}}}, \pairQ x{\bang^\phl{\tilde x} A} 
		}{
			\phl{Q} \seq \set[i]{\pairQ{x_i}{\query \dual{A_i}}}, \br[\tilde x]{\Query}\pairQ z A
		}
	}
\end{displaymath}
\qquad\qquad$\Longrightarrow$ 
\begin{displaymath}\small
	\infer[\tsc{MCutF}]
	{\phl{\res{x\tilde x:\res{\tilde y}Q\br{\tilde P}}(R \pp \tilde R)}
		\cll 
		\Sigma, \tilde\Sigma
	}
	{
		\phl R \cll \thl \Sigma, \phl z:\thl A
		&\quad\ldots\quad
		&
		\phl{Q} \seq \set[i]{\pairQ{x_i}{\query \dual{A_i}}}, \br[\tilde x]{\Query}\pairQ z A
	}
\end{displaymath}

\mypar{Server discarding ($\textsc{w}$).} 
$$
\infer[?]
{\phl{\client xzR} \cll \thl \Sigma, \phl x:\thl {\query A}}
{\phl R \cll \thl \Sigma, \phl y:\thl A}
\qquad
\infer[\textsc{w}]
{\phl R \cll \thl \Sigma, \phl x:\thl{\query A}}
{\phl R \cll \thl \Sigma}
\qquad
\infer[\textsc{c}]
{\phl R \cll \thl \Sigma, \phl x:\thl{\query A}}
{\phl R \cll \thl \Sigma, \phl x:\thl{\query A}, \phl x:\thl{\query A}}
$$

\begin{displaymath}\small
	\infer[\tsc{MCutF}]
	{\phl{\res{x\tilde x:\res{\tilde y}\srv xzQ\br{\tilde P}}(R \pp \tilde R)}
		\cll 
		\Sigma, \set[i]{\query\Sigma_i}
	}
	{
		\infer[\textsc{w}]
		{\phl R \cll \thl \Sigma, \phl x:\thl{\query A}}
		{\phl R \cll \thl \Sigma}
		&
		\quad
		\set[i]{\phl{R_i } \cll \query\Sigma_i, \pairQ{x_i}{\bang A_i}}
		\quad
		&
		\infer[!\ (\tilde u \not= \emptyset)]{
			\srv xzQ \seq \set[i]{\pairQ{x_i}{\query \dual{A_i}}}, \pairQ x{\bang^\phl{\tilde x} A} 
		}{
			\phl{Q} \seq \set[i]{\pairQ{x_i}{\query \dual{A_i}}}, \br[\tilde x]{\Query}\pairQ z A
		}
	}
\end{displaymath}
\qquad\qquad$\Longrightarrow$ 
\begin{displaymath}\small
	\infer[\textsc{w}]{
		\phl{R}
		\cll 
		\Sigma, \set[i]{\query\Sigma_i}
		}{\phl R \cll \thl \Sigma}
\end{displaymath}

\mypar{Server duplicating ($\textsc{c}$).} 
\begin{displaymath}\small
	\infer[\tsc{MCutF}]
	{\phl{\res{x\tilde x:\res{\tilde y}\srv xzQ\br{\tilde P}}(R \pp \tilde R)}
		\cll 
		\Sigma, \set[i]{\query\Sigma_i}
	}
	{
		\infer[\textsc{c}]
		{\phl R \cll \thl \Sigma, \phl x:\thl{\query A}}
		{\phl R \cll \thl \Sigma, \phl x:\thl{\query A}, \phl x:\thl{\query A}}
		&
		\quad
		\set[i]{\phl{R_i } \cll \query\Sigma_i, \pairQ{x_i}{\bang A_i}}
		\quad
		&
		\infer[!\ (\tilde u \not= \emptyset)]{
			\srv xzQ \seq \set[i]{\pairQ{x_i}{\query \dual{A_i}}}, \pairQ x{\bang^\phl{\tilde x} A} 
		}{
			\phl{Q} \seq \set[i]{\pairQ{x_i}{\query \dual{A_i}}}, \br[\tilde x]{\Query}\pairQ z A
		}
	}
\end{displaymath}
\qquad\qquad$\Longrightarrow$ 
\begin{displaymath}\small
	\infer[\textsc c]{
	\phl{\res{x\tilde x:\res{\tilde y}\srv xzQ\br{\tilde P}}(\res{x\tilde x:\res{\tilde y}\srv xzQ\br{\tilde P}}(R \pp \tilde R) \pp \tilde R)}
	\cll 
	\Sigma, \phl x:\thl{\query A}, \set[i]{\query\Sigma_i}
	}{
	\infer[\tsc{MCutF}]
	{\phl{\res{x\tilde x:\res{\tilde y}\srv xzQ\br{\tilde P}}(\res{x\tilde x:\res{\tilde y}\srv xzQ\br{\tilde P}}(R \pp \tilde R) \pp \tilde R)}
		\cll 
		\Sigma, \phl x:\thl{\query A}, \set[i]{\query\Sigma_i}, \set[i]{\query\Sigma_i}
	}
	{
	\infer[\tsc{MCutF}]
	{\phl{\res{x\tilde x:\res{\tilde y}\srv xzQ\br{\tilde P}}(R \pp \tilde R)}
		\cll 
		\Sigma, \phl x:\thl{\query A}, \set[i]{\query\Sigma_i}
	}
	{
		\phl R \cll \thl \Sigma, \phl x:\thl{\query A}, \phl x:\thl{\query A}
		&
		\quad\ldots\quad
		&
		\infer[]{
			\srv xzQ \seq \set[i]{\pairQ{x_i}{\query \dual{A_i}}}, \pairQ x{\bang^\phl{\tilde x} A} 
		}{
			\phl{Q} \seq \set[i]{\pairQ{x_i}{\query \dual{A_i}}}, \br[\tilde x]{\Query}\pairQ z A
		}
	}
		&
		\quad\ldots\quad
		&
		\infer[]{
			\srv xzQ \seq \set[i]{\pairQ{x_i}{\query \dual{A_i}}}, \pairQ x{\bang^\phl{\tilde x} A} 
		}{
			\phl{Q} \seq \set[i]{\pairQ{x_i}{\query \dual{A_i}}}, \br[\tilde x]{\Query}\pairQ z A
		}
	}
}
\end{displaymath}

\section{Example}
\begin{example}[Additive Criss-cross] The process
  $\phl P \colonequals
  \Case{x}{\Case{y}{P_1}{P_2}}{\Case{y}{P_3}{P_4}}$ is a variant of
  multiplicative criss-cross in Example~\ref{ex:criss-cross} that uses
  branching instead. As a derivation:
	\begin{smallequation*}
		\infer=[\with^2]{
			\phl{P}
			\seq
			\pairQ{x}{(\tname\oplus^\phl{y}\tname) \with^\phl{y} (\tcost \oplus^\phl{y}\tcost)}, 
			\pairQ{y}{(\dual\tname \oplus^\phl{x} \dual\tcost)\with^\phl{x}(\dual\tname\oplus^\phl{x}\dual\tcost)}
		}{
			\phl{P_1} \seq \Delta_1
			\qquad&\qquad
			\phl{P_2} \seq \Delta_2
			\qquad&\qquad
			\phl{P_3} \seq \Delta_3
			\qquad&\qquad
			\phl{P_4} \seq \Delta_4
	}
	\end{smallequation*}
	\begin{smallequation*}
		\infer[\oplus_l]{\phl{P_1}  = \inl{x}{\inl{y}{\fwd yx}}
			\seq \Delta_1 =
			\br[y]{\Left}\pairQ{x}{\tname\oplus^\phl{y}\tname}, 
			\br[x]{\Left}\pairQ{y}{\dual\tname \oplus^\phl{x} \dual\tcost}
			}{
				\infer[\oplus_l]{
					\inl{y}{\fwd yx}
					\seq
					\br[y]{\Left}\pairQ{x}{\tname}, 
					\pairQ{y}{\dual\tname \oplus^\phl{x} \dual\tcost}
					}{
					\infer[]{
						\fwd yx
						\seq
						\pairQ{x}{\tname}, 
						\pairQ{y}{\dual\tname}
						}{}
					}
			}
	\end{smallequation*}
	\begin{smallequation*}
		\infer[\oplus_r]{\phl{P_2} = \inr{x}{\inl{y}{\fwd yx}}
			\seq \Delta_2 =
			\br[y]{\Left}\pairQ{x}{\tname\oplus^\phl{y}\tname}, 
			\br[x]{\Right}\pairQ{y}{\dual\tname\oplus^\phl{x}\dual\tcost}
		}{
		\infer[\oplus_l]{
			\inl{y}{\fwd yx}
			\seq
			\br[y]{\Left}\pairQ{x}{\tname}, 
			\pairQ{y}{\dual\tname \oplus^\phl{x} \dual\tcost}
		}{
			\infer[]{
				\fwd yx
				\seq
				\pairQ{x}{\tname}, 
				\pairQ{y}{\dual\tname}
			}{}
		}
	}
	\end{smallequation*}
	\begin{smallequation*}
		\infer[\oplus_l]{\phl{P_3} = \inl{x}{\inr{y}{\fwd yx}}
			\seq \Delta_3 =
			\br[y]{\Right}\pairQ{x}{\tcost \oplus^\phl{y}\tcost}, 
			\br[x]{\Left}\pairQ{y}{\dual\tname \oplus^\phl{x} \dual\tcost}
		}{
		\infer[\oplus_r]{
			\inl{y}{\fwd yx}
			\seq
			\br[y]{\Right}\pairQ{x}{\tcost}, 
			\pairQ{y}{\dual\tname \oplus^\phl{x} \dual\tcost}
		}{
			\infer[]{
				\fwd yx
				\seq
				\pairQ{x}{\tcost}, 
				\pairQ{y}{\dual\tcost}
			}{}
		}
	}
	\end{smallequation*}
	\begin{smallequation*}
		\infer[\oplus_r]{\phl{P_4} = \inr{x}{\inr{y}{\fwd yx}}
			\seq \Delta_4 =
			\br[y]{\Right}\pairQ{x}{\tcost \oplus^\phl{y}\tcost}, 
			\br[x]{\Right}\pairQ{y}{\dual\tname\oplus^\phl{x}\dual\tcost}
		}{
		\infer[\oplus_r]{
			\inl{y}{\fwd yx}
			\seq
			\br[y]{\Right}\pairQ{x}{\tcost}, 
			\pairQ{y}{\dual\tname \oplus^\phl{x} \dual\tcost}
		}{
			\infer[]{
				\fwd yx
				\seq
				\pairQ{x}{\tcost}, 
				\pairQ{y}{\dual\tcost}
			}{}
		}
	}
	\end{smallequation*}
      \end{example}

\end{document}